\documentclass[useAMS,usenatbib]{mn2e} 
\usepackage{epsfig}

\def\spose#1{\hbox to 0pt{#1\hss}} 
\def\lta{\mathrel{\spose{\lower 3pt\hbox{$\mathchar"218$}}      
     \raise 2.0pt\hbox{$\mathchar"13C$}}}      
\def\gta{\mathrel{\spose{\lower 3pt\hbox{$\mathchar"218$}}      
     \raise 2.0pt\hbox{$\mathchar"13E$}}}      
   
\title[Modeling the chemical evolution of $\omega$ Centauri]
{Modeling the chemical evolution of $\omega$ Centauri using
three-dimensional hydrodynamical simulations\thanks{Research
undertaken as part of the Commonwealth Cosmology Initiative
(CCI:www.thecci.org).}}
   
\author[Marcolini et al.]   
       {A. Marcolini$^{1}$, A. Sollima$^{2}$, A. D'Ercole$^{3}$, 
        B.K Gibson$^{1,4}$ and F. R. Ferraro$^{2}$ \\
        $^1$ Centre for Astrophysics, University of Central Lancashire,
        Preston, Lancashire, PR1 2HE, United Kingdom \\
        $^2$ Dipartimento di Astronomia, Universit\`a di Bologna,   
        via Ranzani 1, 40127 Bologna, Italy \\   
        $^3$ Osservatorio Astronomico di Bologna,   
        via Ranzani 1, 40127 Bologna, Italy \\
        $^4$ School of Physics, University of Sydney,
        NSW, 2006, Australia \\}

\date{Accepted ..., Received ...; in original ...}     
   
\pagerange{\pageref{firstpage}--\pageref{lastpage}}     
\pubyear{2004}     
     
\begin{document}     
  
\maketitle     
     
\label{firstpage}     
     
\begin{abstract}    
We present a hydrodynamical and chemical model for the globular
cluster $\omega$~Cen, under the assumption that it is the remnant of
an ancient dwarf spheroidal galaxy (dSph), the bulk of which was
disrupted and accreted by our Galaxy $\sim$10~Gyr ago. We highlight
the very different roles played by Type~II and Type~Ia supernovae
(SNe) in the chemical enrichment of the inner regions of the putative
parent dSph.  While the SNe~II pollute the interstellar medium rather
uniformly, the SNe~Ia ejecta may remain confined inside dense pockets
of gas as long as succesive SNe~II explosions spread them out. Stars
forming in such pockets have lower $\alpha$-to-iron ratios than the
stars forming elsewhere. Owing to the inhomogeneous pollution by
SNe~Ia, the metal distribution of the stars in the central region
differs substantially from that of the main population of the dwarf
galaxy, and resembles that observed in $\omega$~Cen. This
inhomogeneous mixing is also responsible for a radial segregation of
iron-rich stars with depleted [$\alpha$/Fe] ratios, as observed in
some dSphs. Assuming a star formation history of $\sim$1.5~Gyr, our
model succeeds in reproducing both the iron and calcium distributions
observed in $\omega$~Cen and the main features observed in the
empirical $\alpha$/Fe versus Fe/H plane. Finally, our model reproduces
the overall spread of the color-magnitude diagram, but fails in
reproducing the morphology of the SGB-a and the double morphology of
the main sequence.  However, the inhomogeneous pollution reduces (but
does not eliminate) the need for a significantly enhanced helium
abundance to explain the anomalous position of the blue main
sequence. Further models taking into account the dynamical interaction
of the parent dwarf galaxy with the Milky Way and the effect of AGB
pollution will be required.
\end{abstract}    
 
    
\section{Introduction} 
\label{sec:introduction} 
The stellar system $\omega$~Cen (NGC~5139) is unique amongst Galactic
star clusters in terms of its structure, kinematics, and stellar
content. It is the only known globular cluster (GC) showing a clear
[Fe/H] spread \citep[][and references therein]{norris1996}. Recent
photometric surveys have revealed the presence of multiple sequences
in its color-magnitude diagram (CMD), indicating a complex star
formation history \citep{cannon1973,rey2004, sollima2005a}. On the
basis of the analysis of the red giant branch (RGB) morphology, three
different stellar components have been identified: a dominant
metal-poor population ([Fe/H]$\sim -1.6$), an intermediate population
spanning the metallicity range $-1.3<$[Fe/H]$<-1.0$, and a metal rich
component with [Fe/H]$\sim -0.6$ \citep{pancino2000}.

Peculiarities have also been found along the Main Sequence (MS) of the
cluster where an additional blue MS (bMS, comprising $\sim 25\%$ of
the cluster's MS stars) running parallel to the dominant sequence
(rMS) has been resolved \citep{anderson2002, bedin2004}. In contrast
with that predicted by stellar models with canonical chemical
abundances, bMS stars show a metallicity a factor of two higher than
that of the rMS \citep{piotto2005}. A large helium overabundance
($\Delta Y \sim 0.15$) in the bMS sequence has been proposed to
explain its anomalous position in the CMD \citep{norris2004,
piotto2005, sollima2007}.  However, such a large helium abundance
spread poses serious problems for the overall interpretation of the
chemical enrichment history of this stellar system. In fact, no
chemical enrichment mechanism is able to produce the huge amount of
helium required to reproduce the observed MS morphology without
dramatically impacting upon the metal abundance
\citep[cf.][]{karakas2006,sollima2007,romano2007}.

Interestingly, \citet{pancino2002} and \citet{origlia2003} found that
while the metal-poor and intermediate-metallicity stellar populations
of $\omega$~Cen have the expected $\alpha$-element overabundance
observed in halo and GC stars $\langle$[$\alpha$/Fe]$\rangle$ $\simeq$
0.3 \citep{edvardsson1993}, the most metal-rich population
([Fe/H]$\sim -0.6$) shows a significantly lower $\alpha$-enhancement
($\langle$[$\alpha$/Fe]$\rangle$=0.1).  The chemical composition of
the former populations requires that at least part of the gas released
in the interstellar medium (ISM) by SNe~II must have been retained by
$\omega$~Cen over a relatively short ($<$1 Gyr) time
interval. However, the presence of stars with lower values of
[$\alpha$/Fe] at larger values of [Fe/H] indicates SNe~Ia have also
likely contributed to the chemical evolution of $\omega$~Cen.  Since
SNe~Ia explosions occur over a timescale longer than that of SNe~II,
the ratio [$\alpha$/Fe] of the polluted ISM tends to decrease with
time as [Fe/H] increases. Although the chemical properties of the
stellar populations of $\omega$~Cen indicate a prolonged star
formation, there is no consensus as to its duration. An age spread of
3-5 Gyr has been claimed to explain the overall trend of
$\alpha$-elements with metallicity \citep{romano2007}. A somewhat
shorter ($<$ 1.5 Gyr) age difference has been suggested by
\citet{sollima2005b}, comparing theoretical isochrones with the
location of a sample of sub-giant stars with known metallicity.

In any case, the body of evidence collected so far leads to the
hypothesis that $\omega$~Cen was formerly a larger stellar system,
possibly a dwarf galaxy \citep{hughes2000, dinescu1999, majewski2000,
smith2000, gnedin2002, bekki2006, romano2007}, that lost most of its
stars and gas in the interaction with the Milky Way. Under this
hypothesis, self-consistent dynamical models, and coupled N-body +
hydrodynamical simulations \citep{carraro2000, bekki2003,
tsuchiya2004}, succeed in reproducing the main observable
characteristics of this system, assuming a total mass for the initial
object in the range $10^8$-$10^9$ M$_{\odot}$.  Tidal interaction with
the Galaxy, as well as ram pressure stripping by its halo gas, have
also been invoked to explain the stellar structure and kinematics of
the local dwarf spheroidal galaxies (dSphs), and their lack of gas
\citep[e.g.][]{vandenbergh1994,mayer2006}. In this scenario
$\omega$~Cen could retain the ejecta of previous generations of stars
and thereby self-enrich throughout its star formation history (SFH).

We propose a model for $\omega$ Cen based upon 3D hydrodynamical
simulations of an evolving dwarf spheroidal galaxy
\citep[see][]{marcolini2006}.  In Section~2, we describe the evolution
of $\omega$~Cen in the general frame proposed by
\citet{marcolini2006}. The general results and the possibility of an
extended SFH are discussed in Section~3. Section~4 is devoted to the
simulation of the systems' CMDs, while our conclusions are drawn in
Section~5.

\section{The model}   
\label{sec:model} 
\subsection{Qualitative framework}
\label{sec:qualitative_framework}

To fully appreciate the chemical enrichment history of $\omega$~Cen it
is necessary to understand both its SFH and the associated role of SNe
feedback on its dynamical and chemical
evolution. \citet{marcolini2006} provide a useful framework in which
to pursue work of this nature, in that their three-dimensional
hydrodynamical simulations of isolated dSphs are directly comparable
to the objects from which $\omega$~Cen has been proposed to have
originated.  In this study, the authors assumed a prolonged
intermittent SFH, tracking the roles of both SNe~Ia and SNe~II in the
chemical enrichment history of the system. These models, although
intended to give a general picture of dSphs, were tailored to the
Draco dwarf galaxy.

The total energy released by the SNe~II explosions is much larger than
the binding energy of the gas. However, efficient radiative losses
enables the galaxy to retain most of its gas, providing the necessary
fuel source for the aforementioned prolonged SFH.

The SNe~II are more concentrated toward the galactic centre, where
their remnants overlap forming a single cavity composed of a network
of tunnels filled by hot rarefied gas. The bulk of the ISM is pushed
outward to the edge of the cavity.  Once the SNe~II cease exploding
($\sim$30~Myr after each star burst episode), the global cavity
collapses and the ISM flows back down the potential well, giving rise
to a new starburst (see Fig.~3 in Marcolini et al. 2006).  In
conclusion, the gas content in the centre has an oscillating temporal
profile whose period is given by the time interval between two
successive starbursts.

Given their lower rate, SNe~Ia do not significantly affect the general
hydrodynamical behaviour of the ISM, but their role is relevant for
the chemical evolution of the stars. Because of the longer timescales
over which they contribute, the SNe~Ia progenitors produced in
previous starbursts continue to explode during the quiescent periods,
even after the gas has settled back to the central region. The higher
ambient gas density (together with the lower explosion rate) results
in SNe~Ia remnants being isolated from one another during this phase
(see Fig.~14 in Marcolini et al. 2006).
As a consequence of this inhomogeneous pollution, stars forming in the
regions occupied by SN~Ia remnants (hereafter referred to as ``SNe Ia
pockets'') have lower [$\alpha$/Fe] ratios and higher [Fe/H] ratios
than those formed elsewhere. This effect is particularly important for
the chemical evolution of the central region, where the SN~Ia rate is
greater and where SN~Ia pockets formed in the outer zones are returned
to the galactic centre, with the gas flow during the re-collapse
phases.  The cycle of expansion and re-collapse experienced by the
galactic gas tends to homogenise the ISM and SNe ejecta rather rapidly
($\sim$10$^8$~yr).

These authors also found that only $\sim 20\%$ of the metals ejected
by SNe were present in the region where the stars were assumed to
form, despite the absence of a galactic wind. The missing metals were
pushed to larger distances by the continuous action of the SN
explosions and did not enrich the forming stars (thus mimicking an
``outflow'').

\begin{figure*}    
\begin{center}    
\psfig{figure=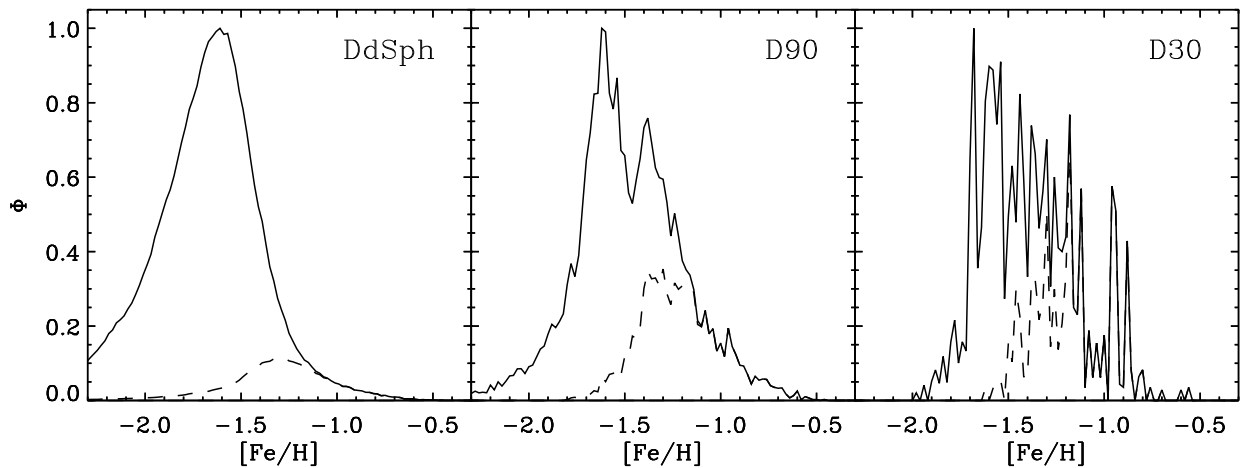}    
\end{center}   
\caption{[Fe/H] distribution function of the long lived stars for the
whole ``entire dSph volume'' (left panel; $r_{\star, \rm c}=130$ and
$r_{\star, \rm t}=650$ pc), and two more centrally located regions
called ``central region'' (middle panel, $r_{\star, \rm t}=90$ pc) and
``nuclear region'' (right panel; $r_{\star, \rm t}=30$).  The dashed
lines represent stars with [$\alpha$/Fe]$\le 0.2$. See text for more
details.}
\label{fig:zstelle} 
\end{figure*}

\subsection{SNe chemical abundances}
\label{sec:sneaboundances}

The numerical method used to treat the SNe explosions is described in
\citet{marcolini2006}. Here we describe briefly the nucleosynthesis
prescription assumed in this paper. Nucleosynthesis products of SNe~II
and SNe~Ia are taken from \citet[][and references
therein]{iwamoto1999}. Their Table 3 summarises the mean SN~II yields
averaged over the $10-50~$M$_{\odot}$ precursors, and the yields for
different models of SN~Ia (we will refer to their W7 model). We
calculate the $\alpha$-element enrichment by summing the different
contributions of O, Mg, Ne, Si, S, Ca, Ti and Ar. We find that a
slightly enhanced Ca production (50\%) over the mean SNe~II value 
helps to better
reconcile model predictions with observations. Note that this increase
lies within the uncertainties associated with the different SNe
models. These yields depend slightly on metallicity, unless metals are
completely absent \citep{woosley1995,limongi2000}; we do not consider
this dependence because it does not affect the [$\alpha$/Fe] features
of the stars with [Fe/H] $> -3$ \citep{Goswami2000}. The adopted
yields are listed in Table~1, together with the adopted solar
abundances \citep[adopted from][]{grevesse1998}.

\begin{table} 
\centering 
\begin{minipage}{80mm} 
\caption{Assumed average SNe yields in solar masses$^a$} 
\label{}   
\begin{tabular} {|l|c|c|c|}
\hline  
        & SN II$^b$ & SN Ia$^c$ & Sun$^d$\\
species &  \\ 
 
\hline

$M_{\rm tot,ej}$  & 17.0 & 1.4 &  \\
$M_{\rm Z}  $     &  3.0 & 1.4 &  0.016 \\
$M_{\rm Fe} $     & $9.07\times 10^{-2}$  & 0.75 & 1.23 $\times10^{-3}$ \\
$M_{\rm O}  $     & 1.80 & 0.14 &  7.56 $\times10^{-3}$\\
$M_{\rm Mg} $     & 0.12 & $8.57 \times 10^{-3}$ & 6.45 $\times10^{-4}$ \\
$M_{\rm Si} $     & 0.12 & 0.16 &  6.96 $\times10^{-4}$  \\
$M_{\rm Ca} $     & $8.80\times 10^{-3}$ & $1.19\times 10^{-2}$ & 
                                             6.41 $\times10^{-5}$ \\
$M_{\alpha} $    &  2.34 & 0.43 & 1.12 $\times10^{-2}$   \\
$[\alpha/$Fe]  & 0.45 &$-1.20$ & 0\\

\hline
\end{tabular}   
\par  
\medskip
$^a$ From Iwamoto et al. 1999 \\
$^b$ Averaged over the progenitor mass range 10-50~M$_{\odot}$ \\ 
$^c$ Model W7 \\
$^d$ Solar abundances from \citet{grevesse1998}.
\end{minipage}
 \label{tab:models}   
\end{table}

\subsection{Color-magnitude diagrams}    
\label{sec:colormagnitude}  

As a first step, in order to simulate the morphology of the CMD we use
the evolutionary tracks of \citet{cassisi2004} and
\citet{pietrinferni2006}, calculated for two different
$\alpha$-enhancement levels ([$\alpha$/Fe]$=0.0$ and
[$\alpha$/Fe]=$+$0.4).  As we will show in Sec.~\ref{sec:results} a
number of stars in our models show [$\alpha$/Fe] values that lie
outside these limits, especially at sub-solar values of
[$\alpha$/Fe]. In order to take these stars into account, we
extrapolated for each metallicity $Z$ the evolutionary tracks down to
values of [$\alpha$/Fe]=$-$0.4.

\begin{figure*}    
\begin{center}    
\psfig{figure=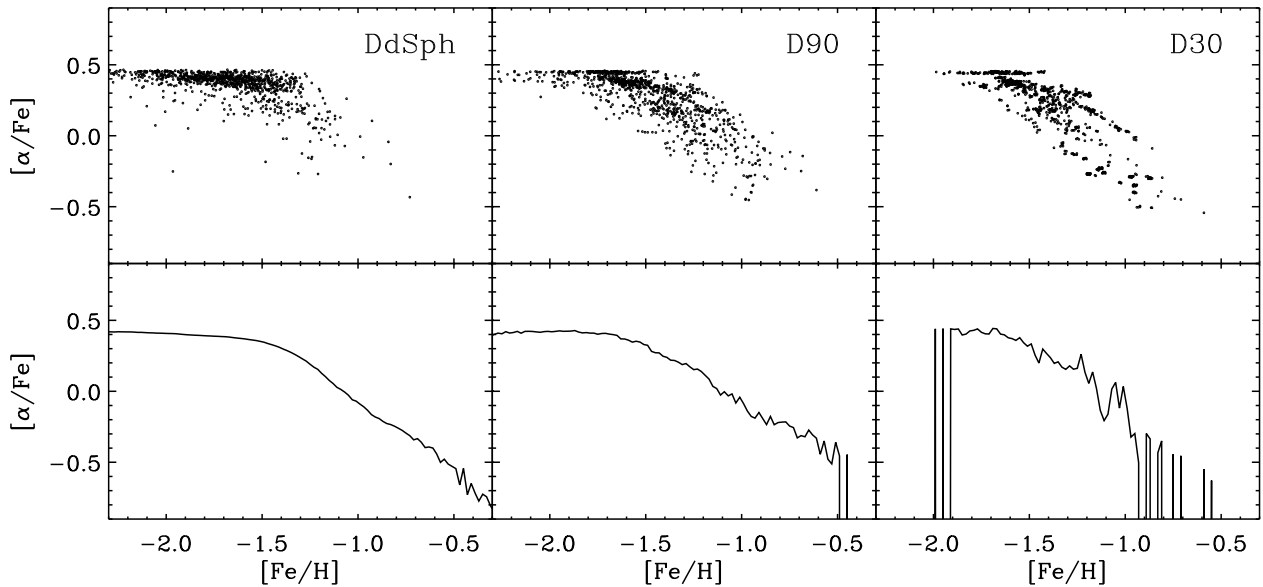}    
\end{center}   
\caption{Upper panels: abundance ratio [$\alpha$/Fe] plotted against
[Fe/H] of $N_{\rm S}=1000$ sampled stars for the three different
regions as in Fig.1. Lower panels: mean values of [$\alpha$/Fe] for
the same stellar sampling.}
\label{fig:zstelleoss} 
\end{figure*}

\begin{figure*}    
\begin{center}    
\psfig{figure=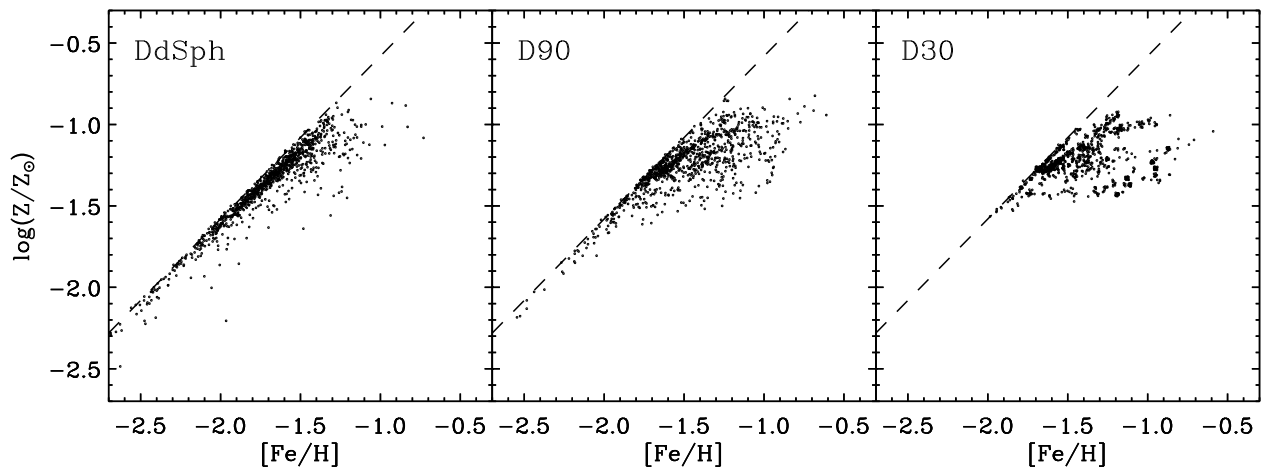}    
\end{center}   
\caption{[Fe/H] of $N_{\rm S}=1000$ sampled stars plotted against the
metallicity Z for the three different regions as in
Fig.~\ref{fig:zstelle}. The dashed line represents the linear relation
given by eq.~(2) in the case of pure SN II enrichment.}
\label{fig:feh-z} 
\end{figure*}

For each star, our code stores its iron abundance [Fe/H], its
metallicity $Z$, its $\alpha$-enhancement [$\alpha$/Fe], and the epoch
of formation.  Then, for each star we randomly extracted 100 masses
from a Salpeter (1955) IMF in the mass range 0.5 M$_{\odot}<$M$<0.9$
M$_{\odot}$ and placed them in the theoretical $\log(L/L_{\odot})$,
$\log T_{\rm eff}$ plane by means of suitable interpolations of the
adopted evolutionary tracks. Luminosities and effective temperatures
were transformed into the desired photometric system by interpolation
within appropriate tables for photometric conversions (for a detailed
description see Cassisi et al. 2004). Following these prescriptions we
simulated a CMD of 100,000 stars in the Johnson-Cousin and in the ACS
VEGAMAG photometric systems.

In section~\ref{sec:comparing_cmd} we will compar our synthetic CMD
with the observed CMDs by \citet{sollima2005a}, \citet{ferraro2004}
and \citet{sollima2007}. To convert the absolute magnitudes into the
apparent ones, we need to assume a distance modulus and a reddening
correction. In the following, we adopt $(m-M)_0=13.70$
\citep{bellazzini2004, delprincipe2006}, the reddening E(B-V)=0.11
\citep{lub2002}, the extinction coefficients $A_{B}=4.1~E(B-V)$,
$A_{V}=3.1~E(B-V)$, $A_{R}=2.32~E(B-V)$ \citep{savage1979} for the
Johnson-Cousin passbands and $A_{F435W}=4.02~E(B-V)$,
$A_{F625W}=2.62~E(B-V)$ \citep{sirianni2005} for the ACS
passbands. For each star a random photometric error extracted from a
Gaussian distribution with $\sigma$ equal to the average photometric
error estimated in that magnitude range by the above authors has been
added.

\begin{figure*}    
\begin{center}    
\psfig{figure=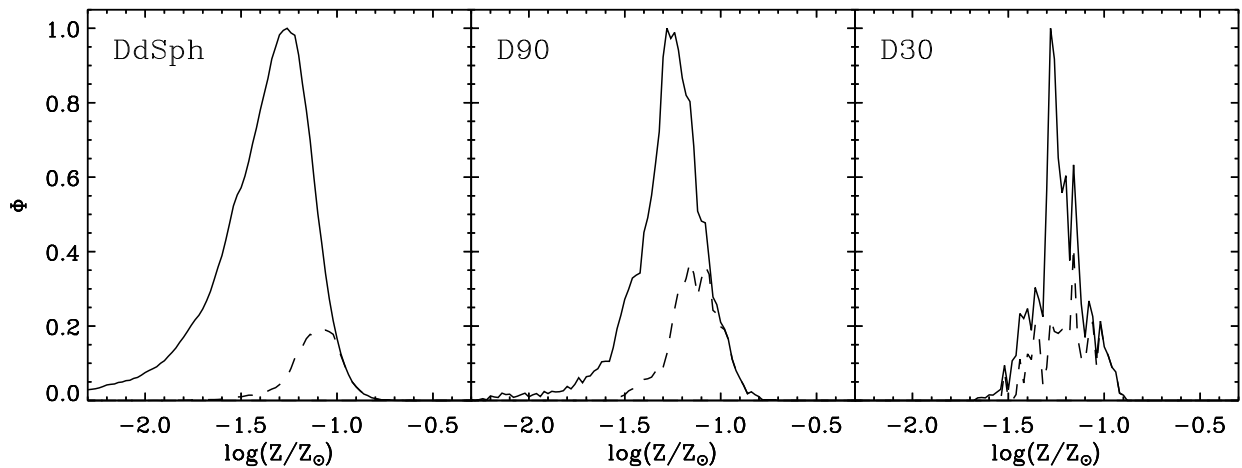}    
\end{center}   
\caption{Metallicity ($Z$) distribution function of the long lived
stars for the ``entire dSph volume'' (left panel), the ``central region''
(middle panel) and the ``nuclear region'' as in
Fig.~\ref{fig:zstelle}. The dashed lines represent stars with
[Fe/H]$\ge-1.4$.}
\label{fig:zstellez} 
\end{figure*}

\subsection{Adapting the model to $\omega$ Centauri}
\label{sec:adaptingmodel}

The 3D simulations employed here were generated using the Marcolini
et~al. (2006) hydrodynamical code.  As pointed out in
Section~\ref{sec:introduction}, in the scenario that assumes
$\omega$~Cen is the remnant of a dwarf galaxy accreted by the Milky
Way, most of the initial mass of the proto-galaxy has been stripped,
presumably by the tidal interaction with the Milky Way. Unfortunately,
our model cannot account for the self-consistent interaction of the
proto-galaxy with the Milky Way.  Thus, we adopt a model originally
tailored to the Draco dSph \citep[see][] {marcolini2006}, but now
focusing our analysis on the central regions of the models - i.e.,
those regions expected (potentially) to survive as a ``globular
cluster'' like $\omega$~Cen.

In the model, the SFH is represented as a sequence of instantaneous star
bursts separated by quiescent periods ($\Delta t=60$ Myr). At the end
of the entire star-formation process, nearly $6\times 10^5$ M$_{\odot}$
of stars have been formed, and $6\times 10^3$ SNe~II exploded.  The
Draco galaxy has a stellar mass content smaller by a factor of 4-8
\citep[e.g.][]{meylan1995, vandeven2006, mateo1998} and a much more
extended stellar distribution than $\omega$~Cen. These differences are
possibly even more marked if we consider the structure of $\omega$~Cen
{\it before} the interaction with the Milky Way.  Thus, in order to
match the chemical properties of the dominant population of the
system, the number of SNe and the initial amount of gas in the
``proto'' $\omega$ Cen have been scaled appropriately, and the star
formation suppressed after $\sim$1~Gyr.

$\omega$~Cen has a stellar mass content of $\sim 3 \times 10^6$
M$_{\odot}$ \citet{vandeven2006} and a radial profile consistent
with a King model with $r_{\rm c}=4.1$~pc and $r_{\rm t}=83$
pc \citep{ferraro2006}. We followed the chemical evolution of the
stellar population of the dSph progenitor in three different regions,,
defined as follow: $i)$ the ``entire dSph volume''
with a tidal radius $r_{\star, \rm t}=650$ pc (see Marcolini et
al. 2006 for more details); $ii)$ the ``central region'' within a
radius $r_{\star, \rm t}=90$ pc; and, $iii)$ the ``nuclear region'', with
$r_{\star, \rm t}=30$ pc. These latter two regions are dense enough to
survive the tidal disruption incurred through the interaction with the Galaxy,
and appear today like a globular cluster. We thus focus our analysis
on the central and nuclear regions, and will refer to them as models
D90 and D30, respectively.

\section{Results}
\label{sec:results}

Figure~\ref{fig:zstelle} illustrates the [Fe/H] distribution of the
long-lived stars (with masses M $<$ 0.9 M$_{\odot}$) formed over
1~Gyr in the three regions described above.  The distribution shown
in the left panel is a rather smooth log normal-type distribution,
similar to those found in dSphs \citep[e.g.][]{bellazzini2002,
babusiaux2005, koch2006, battaglia2006, bosler2007}. The profile in the
``rentral region'' (middle panel), though, shows a bimodal structure
similar to that observed in $\omega$~Cen \citep[e.g.][]{norris1996,
suntzeff1996}, with a maximum at [Fe/H]=$-1.6$ and a secondary peak at
[Fe/H]$\sim-1.3$ accounting for $\sim25$\% of the cluster's stellar
content. The different shape of the latter's distribution is due to
the higher SN~Ia rate in this region, as well as to the fact that
external iron rich SN~Ia pockets converge toward the centre during the
re-collapsing stages of the ISM evolution.  The right panel in
Fig.~\ref{fig:zstelle} shows the distribution in the ``nuclear
region''.  The noise present is due to sampling statistics and to the
fact that the volume considered is comparable to that of single
pockets.  In each panel only a negligible fraction ($<1$\%) of stars has a
metallicity [Fe/H]$>-0.7$.

The upper panels of Fig.~\ref{fig:zstelleoss} show the
[$\alpha$/Fe]-[Fe/H] plane for $N_{\rm s}=1000$ sampled stars. As in
Fig.~\ref{fig:zstelle}, the left panel refers to the whole galaxy
while the central and the right panels refer to the central and the
nuclear regions, respectively. The lower panels represent instead the
mean value of [$\alpha$/Fe]. Because of the SN~Ia contribution, the
Fe-rich stars are, as expected, $\alpha$-depleted: we find mean values of
$\langle [\alpha$/Fe$] \rangle \sim 0.4$ at [Fe/H]$=-1.7$ and $\langle
[\alpha$/Fe$] \rangle \sim 0.2$ at [Fe/H]$=-1.3$, with a large spread
in the [$\alpha$/Fe] distribution at [Fe/H]$>-1.5$ .

In general, our mean value of [$\alpha$/Fe] in the high metallicity
regime (Fe/H$> -1.0$) is slightly smaller than that observed by
\citet{smith2000} and \citet{origlia2003}. This discrepancy is most
likely due to the simple form of the SFH we assumed.  In
Section~\ref{sec:longersf} we will show that this discrepancy, as well
as the lack of metal-rich ([Fe/H]$>-0.7$) stars, may be reduced by
adopting a more protracted SFH. Indeed, assuming that the star
formation can last beyond 1~Gyr at a lower rate, high-metallicity
stars ([Fe/H]$\gta-1.3$) may continue to form with high [$\alpha$/Fe]
values because of the freshly-expelled ejecta from recent SNe~II
explosions.

\begin{figure}    
\psfig{figure=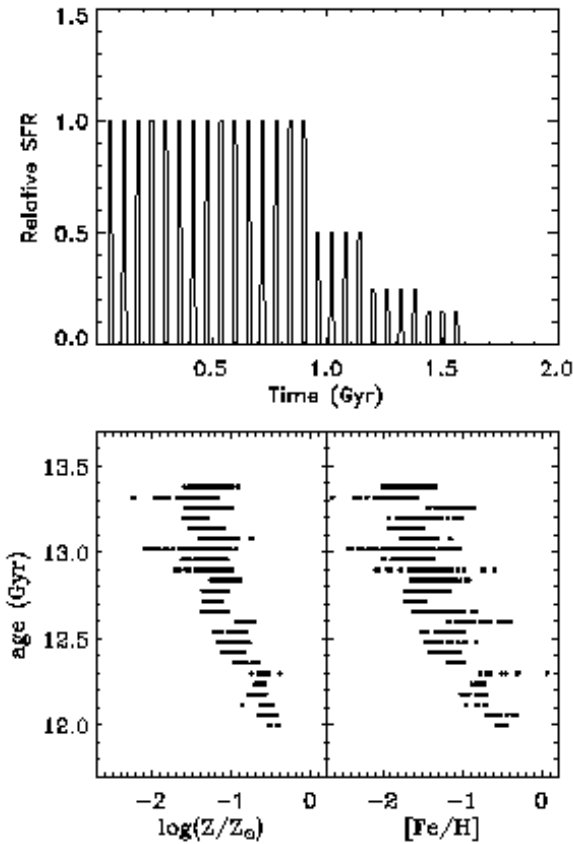,width=0.45\textwidth}    
\caption{Upper panel: temporal profile of tha assumed star formation
rate in the LONG model. Lower panel: age-$Z$ and age-[Fe/H]
distributions of 1000 sampled stars for the D90LONG model. Note the
metallicity spread among coeval stars.}
\label{fig:age} 
\end{figure}

We must re-iterate the difference between the pollution of iron and
that of metals, as a whole.  It is clear from Table~1 that the ratio
between the iron mass and the mass of all the metals is rather
different in the ejecta of the two types of SNe. Thus, when the
contribution of the SNe~Ia to the ISM pollution becomes relevant, the
metal mass fraction $Z$ is not linearly proportional to the amount of
iron [Fe/H]. It can be shown that, with our assumptions, it is:
\begin{equation}
Z=0.015\times 10^{[Fe/H]}10^{[\alpha/Fe]}
\left [1+0.08\left(2.82\times10^{-[\alpha/Fe]}-1\right)\right].
\end{equation}

Figure~\ref{fig:feh-z} shows the $\log(Z/Z_{\odot})$-[Fe/H] diagram
for the same sample of stars of Fig.~\ref{fig:zstelleoss}. The dashed
line represents the $\log(Z/Z_{\odot})$-[Fe/H] relation for stars
formed in regions enriched only by SNe~II
\footnote{This relation is obtained from the previous formula assuming
[$\alpha$/Fe]=$+$0.45, representative of SNe~II ejecta
(cf. Table~1)}:

\begin{equation}
\log(Z/Z_{\odot})=0.42+[Fe/H],
\end{equation}
assuming $Z_{\odot}=0.016$   \citep{grevesse1998}.

The $Z$ distribution for the long-lived stars is plotted in
Figure~\ref{fig:zstellez}. The bimodal structure of the [Fe/H]
distribution (see Fig.~\ref{fig:zstelle}) is now absent and the spread
of the $Z$ distribution is similar to the spread of the dominant
[Fe/H] poor population. In fact, the number of SNe~Ia occurring after
1~Gyr accounts for $\sim 20$\% of the iron content but only $\sim
1.5$\% of the produced metal mass. Therefore, in the regions polluted
inhomogeneously by SNe~Ia, the [Fe/H] ratio increases dramatically,
but the metal mass fraction $Z$ remains basically unaffected
(c.f. Figure~\ref{fig:feh-z}). This effect is particularly important
in the central regions, where the stars formed in ``SNIa pockets'' can
constitute $\sim$30\% of the entire cluster population. In
Figure~\ref{fig:zstellez} the $Z$ distribution for stars with
different iron content is shown for the three regions described in
Section~\ref{sec:adaptingmodel}. Note that, in the central regions,
Fe-rich ([Fe/H]$\gta -1.4$) and Fe-poor stars span nearly the same $Z$
range. As we will see in Section~\ref{sec:comparing_cmd}, this
difference influences the position of the stars in the CMD.

\subsection{Longer star formation history}   
\label{sec:longersf} 

As discussed in the previous section only a negligible fraction of stars
in our model have a [Fe/H]$>-0.7$ and the mean value of [$\alpha$/Fe] 
in the high metallicity regime ([Fe/H]$>-1.0$) is slightly smaller
than that observed. 
These discrepancies can be traced to the
simple form of the SFH adopted.  
Indeed, in Section~\ref{sec:adaptingmodel}, we assumed that the
star formation of our reference model is suddenly halted after
$\sim$1~Gyr, due to the interaction with the Milky Way. However, the
effect of ram pressure stripping and tidal interactions in more
massive dwarf galaxies is likely to be more prolonged, and the time to
completely strip the gas from such a system may be of the order of
several Gyrs, depending upon the strength of the ram pressure and the
orbit of the system \citep[e.g.][]{marcolini2003, mayer2006}.

To explore this possibility, we assumed that after 1~Gyr, the system
still forms stars for a further 0.6~Gyr, but at a lower rate.  This
hypothesis is consistent with the ancient proto-galaxy losing most
(but not all) of the gas during its first interaction with the Milky
Way. This newly adopted SFH is plotted in the upper panel of
Figure~\ref{fig:age}. In the following, we refer to this model as LONG.

The reduction of the ISM reservoir during this 0.6~Gyr period leads to
a corresponding reduction in the rate of the SNe~II, whose progenitors
are short-lived stars. Conversely, in the same period, the SN~Ia rate
does not decrease significantly because, in this case, the precursors
have longer lifetimes and their explosion rate still depends on the
past, higher, SFR. Thus, the rapid gas depletion, together with a
rather unaffected iron production, leads to a rapid metal enrichment
of the stars forming after the stripping episode, during the final
0.6~Gyr.

\begin{figure*}    
\begin{center}    
\psfig{figure=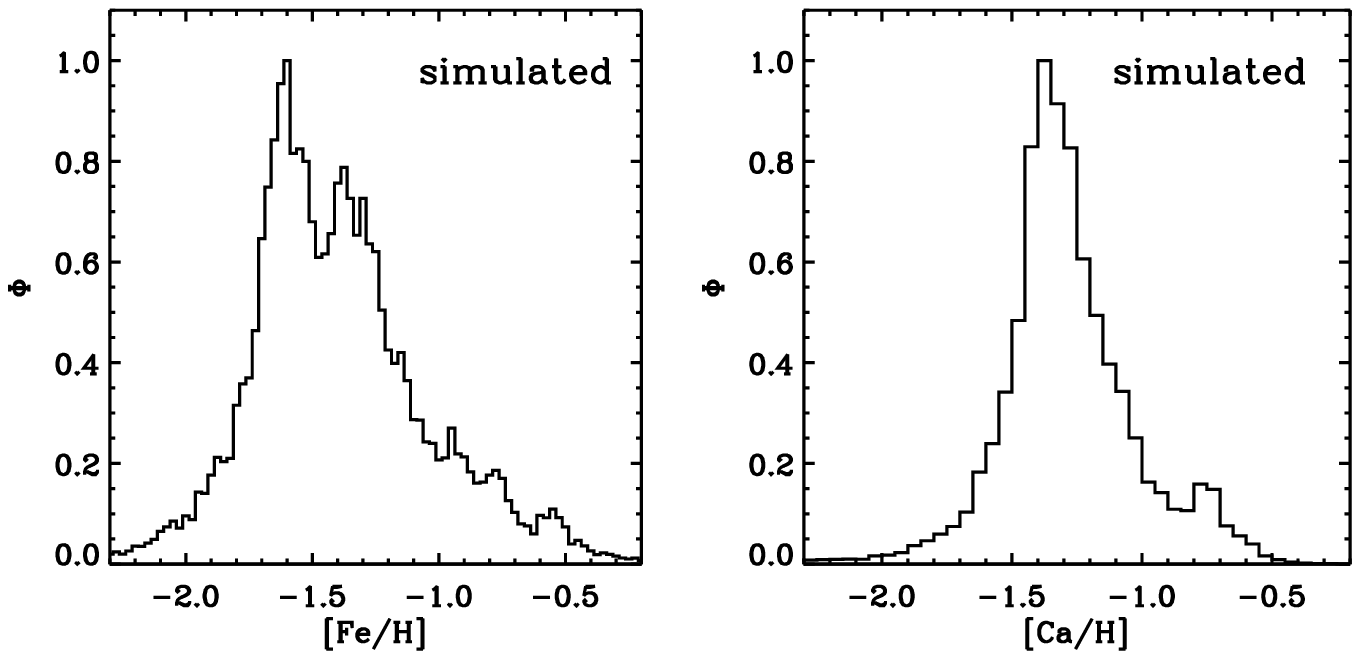,width=0.95\textwidth} 
\\
\psfig{figure=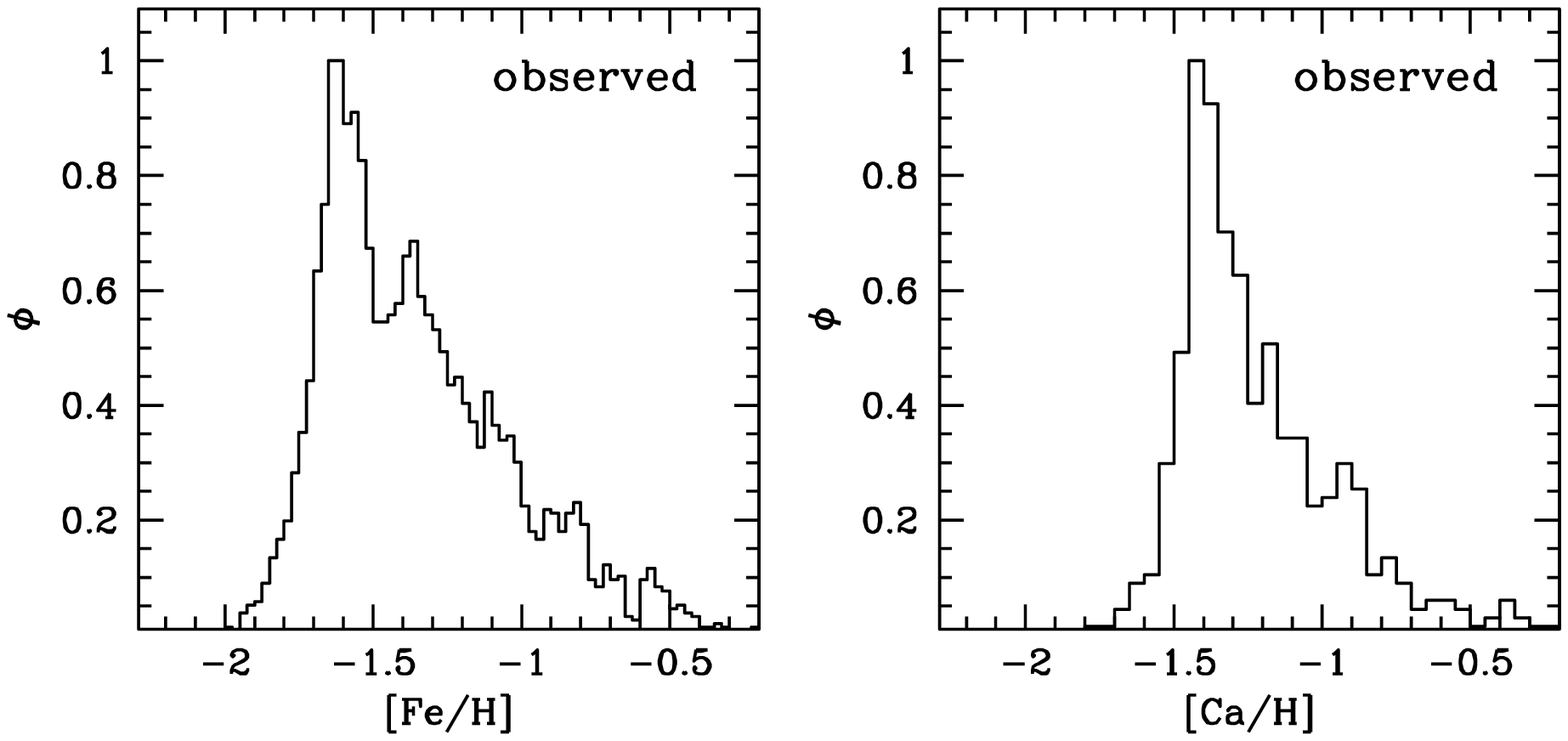,width=0.95\textwidth}
\end{center}   
\caption{Left panels: theoretical (D90LONG model) and observative
  \citep{sollima2005a} [Fe/H] distributions of long lived star (upper
  and lower panels, respectively). Right panels: theoretical and
  observative \citep{norris1996} [Ca/H] distributions of long lived
  stars (upper and lower panels, respectively).}
\label{fig:ztot} 
\end{figure*}

As already discussed in Section~\ref{sec:adaptingmodel}, in our
original model all the starbursts had the same intensity, and the ISM
remains entirely bound to the galaxy. Unfortunately, we cannot mimic
the gas stripping by simply reducing artificially the amount of ISM in
our model; in this way, in fact, the metals mixed with the ISM would
be removed as well and the relative role of SNe II and SNe Ia as
polluters would not change. Instead, the metal enrichment is obtained
by increasing in time the quantity of SN ejecta present in the
simulation; in order to take into account the corresponding drop of
the SNII rate due to the decreasing SF over the last 0.6 Gyr, in
parallel with the rather unaffected SNIa rate (see above), the
relative rates of increasing metal enrichment due to the two SNe types
are chosen a posteriori to best fit the data. Not having strong
constraints on the orbit of first interaction of the proto-galaxy with
the Milky Way (and how the gas stripping proceeded), this remains a
necessary free paramenter.

In the lower panel of Fig.~\ref{fig:age} we plot the age-metallicity
diagram for the D90LONG model. As evident, the metallicity increases
with time even if a large spread is always present, in agreement with
the most recent determinations (e.g. Hilker et al. 2004; Stanford et
al. 2006; Villanova et al. 2007). Note that the spread is more evident
in the [Fe/H] distribution, as opposed to the $Z$ one, because of the
stronger influence of the inhomogeneous SNe~Ia enrichment on the Fe
abundance.  We find that we can generate the metallicity distribution
functions (and associated observational constraints, as noted below)
of $\omega$~Cen assuming a SFH lasting 1.6~Gyr, in agreement with the
timescale proposed by \citet{ferraro2004} and \citet{sollima2005b}.

In Figure~\ref{fig:ztot} we show the [Fe/H] and [Ca/H] distributions
obtained by the D90LONG model, together with those observed.
Comparing the [Fe/H] distribution in the upper-left panel with the
corresponding one in Figure~\ref{fig:zstelle} we note that the shape
of the [Fe/H] distribution remains unaffected at low metallicity
([Fe/H]$<-1.2$), but for [Fe/H]$\gta-1.2$ the D90LONG model predicts a
larger number of stars and a small peak at [Fe/H]$\sim-0.6$. The
agreement with the [Fe/H] distribution inferred by the observations of
\citet{sollima2005a} is now quite good.

In the same figure we also compare the [Ca/H] distribution found by
our model and the one inferred by \citet{norris1996}. Again, the two
distributions are in good agreement. Actually, we tend to overestimate
the number of stars having [Ca/H]$>-0.8$, but as stated by
\citet{norris1996}, their selection criteria tends to underestimate
the number of stars in this range. Note that, contrary to the iron,
the amount of calcium ejected by a single SN~Ia is comparable to that
expelled by a SN II (SN~Ia$_{ej,Ca}$/SN~II$_{ej,Ca}\sim
1.4$)\footnote{Note that this ratio is $\sim 8$ for iron}. Thus, the
two peaks visible in the [Fe/H] distribution become nearly
indistinguishable in the [Ca/H] distribution.

\begin{figure}    
\psfig{figure=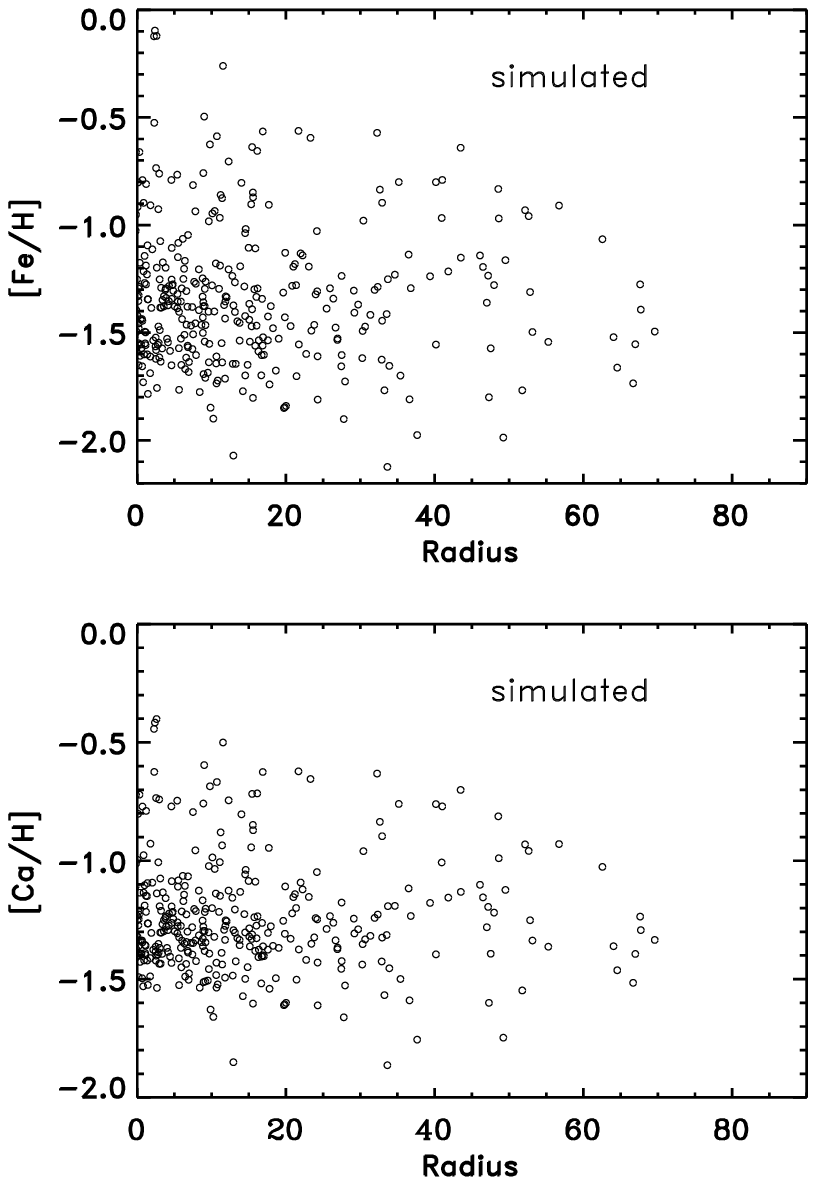,width=0.40\textwidth}
\\
\psfig{figure=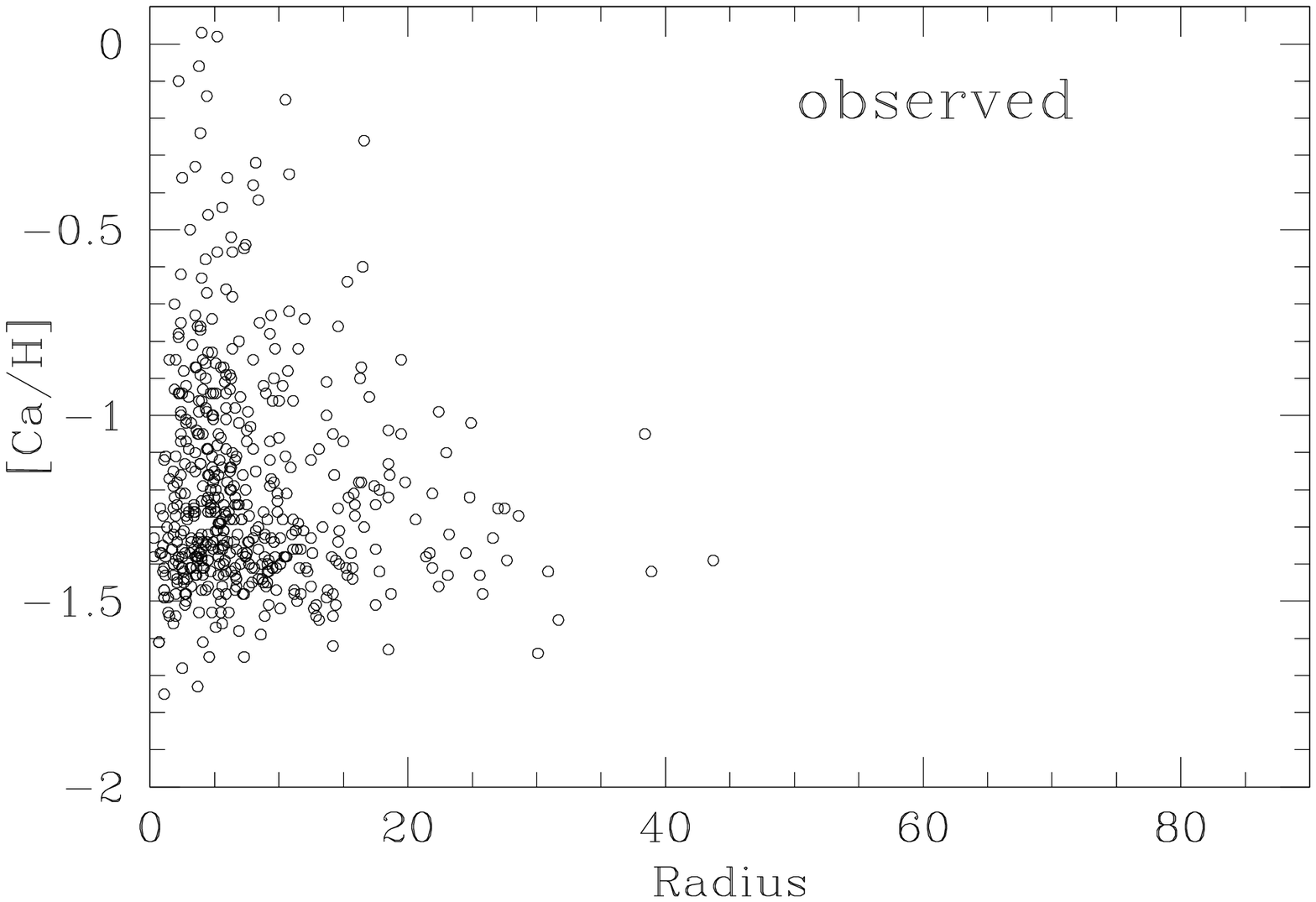,width=0.40\textwidth} 
\caption{[Fe/H] (upper panel) and [Ca/H] (middle panel) projected
radial distributions of 500 sampled stars for the D90LONG model. The
observed [Ca/H] projected radial distributions by Norris et al. (1997)
is shown in the bottom panel. Radius is given in pc.}
\label{fig:zraggio} 
\end{figure}

In Fig.~\ref{fig:zraggio} we plot the projected radial distribution of
[Fe/H] (upper panel) and [Ca/H] (central panel) for a sample of 500
stars. As already discussed, the most iron-rich stars are more
concentrated toward the centre
(c.f. Section~\ref{sec:qualitative_framework}). In the bottom panel,
the [Ca/H] distribution observed by \citet{norris1997} is shown.  We
find good agreement between the observed distribution and that
predicted by the D90LONG model.

\begin{figure*}    
\begin{center}    
\psfig{figure=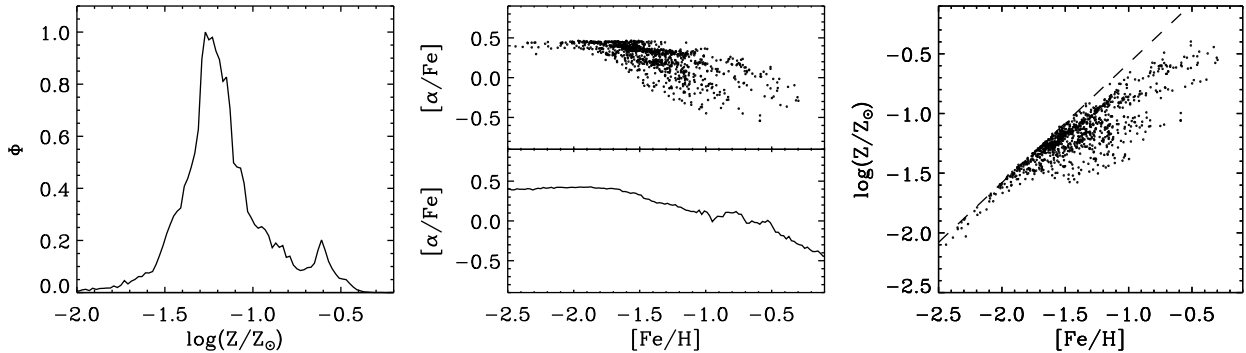,width=0.95\textwidth}
\end{center}   
\caption{ Left panel: log($Z$) stellar distribution.  Central panels:
[$\alpha$/Fe]-[Fe/H] diagram for 1000 sampled stars (top), and
$\langle$[$\alpha$/Fe]$\rangle$-[Fe/H] relation (bottom), where the
brakets indicate the mean value of [$\alpha$/Fe]. Right panel:
log($Z/Z_{\odot})$-[Fe/H] diagram of 1000 stars sampled in the central
region. All the results refer to the model D90LONG}
\label{fig:ztot2} 
\end{figure*}

Figure ~\ref{fig:ztot2} (left panel) illustrates the log($Z$)
distribution predicted by the D90LONG model. The peak at
log($Z/Z_{\odot})=-0.65$ is due to stars born in the final episode of
star formation which have $-0.8\le$[Fe/H]$\le-0.4$. These stars have
ratios [$\alpha$/Fe] in the range and $-0.3\le [\alpha$/Fe$]\le+0.3$,
as shown by the two central panels of Figure~\ref{fig:ztot2}. Comparing
the central panel of this latter figure with the central panels of
Figure~\ref{fig:zstelleoss}, we note that the mean [$\alpha$/Fe]
values at [Fe/H]=$-$1.7, $-$1.3, and $-$0.6 are now
$<[\alpha$/Fe]$>\sim+0.4, +0.2$, and $+0.0$, respectively, in good
agreement with the values inferred by \citet{origlia2003}. The
improvement at higher [Fe/H] is due to the temporal tail of the star
formation. Indeed, as the star formation proceeds (though at a lower
rate), the new SNe~II explosions release fresh $\alpha$-elements which
raise the [$\alpha$/Fe] ratio of most of the newly-formed stars.
Finally, the right panel of Fig.~\ref{fig:ztot2} emphasises the
dispersion of the stellar metal fraction in the log($Z$)-[Fe/H]
diagram.

\begin{figure*}    
\psfig{figure=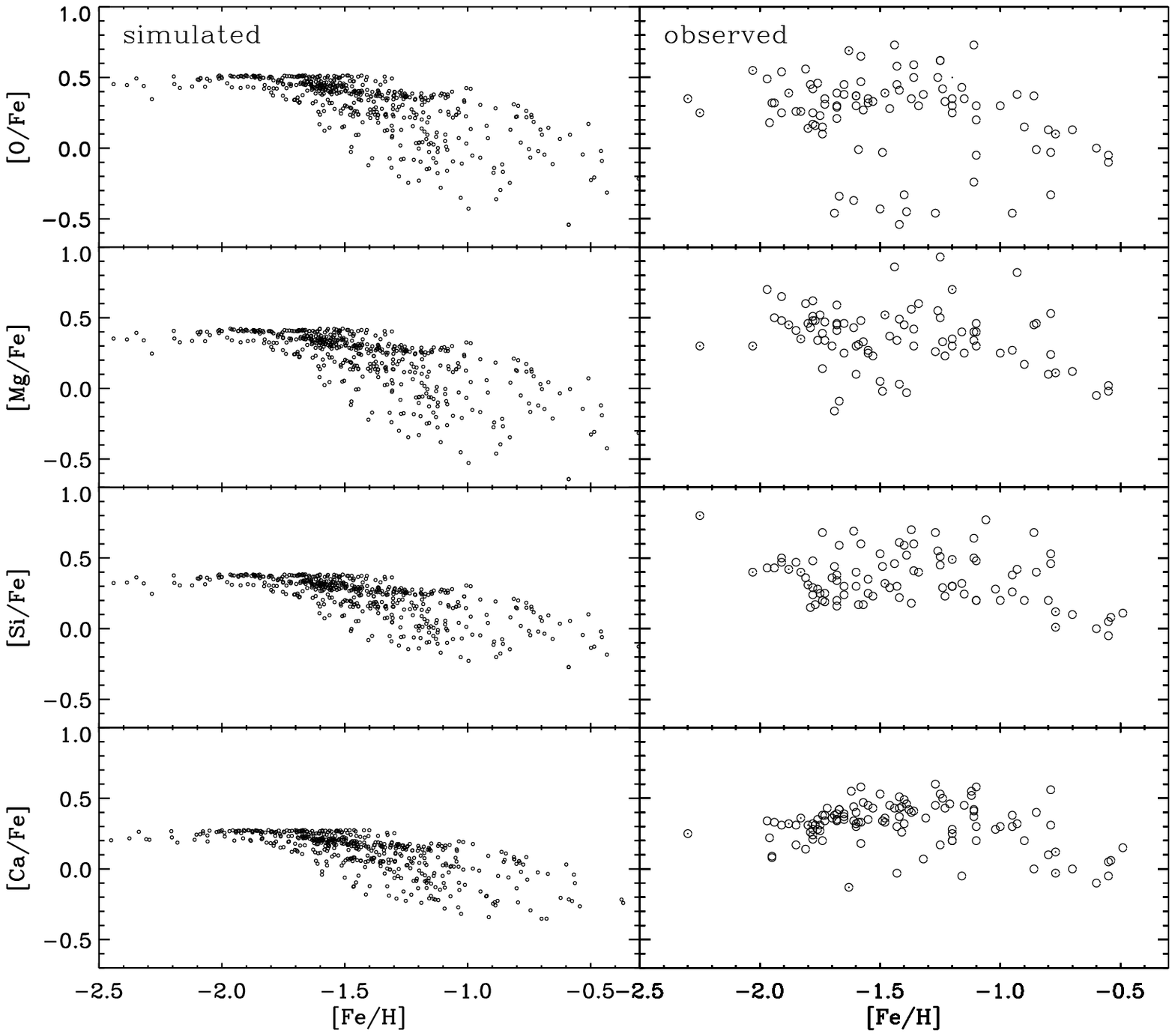,width=0.9\textwidth}    
\caption{Predicted abundance ratios of several $\alpha$-elements
to iron as a function of [Fe/H] of $N_{\rm S}=500$ sampled stars
(left panels). Observative data are in the right panels (see text for
the references).}
\label{fig:ofe} 
\end{figure*}

To better compare our model with observations we plot separately in
Figure~\ref{fig:ofe} the abundance patterns of several
$\alpha$-elements (O, Mg, Si, and Ca) as predicted by our model,
together with a dataset collected from the literature
\citep[]{francois1988, brown1993, norris1995, smith1995, smith2000,
pancino2002, vanture2002, origlia2003, villanova2007}. Altough the
data are in good agreement, discrepancies as large as 0.5~dex, are
occasionally present \citep[for an extensive discussion of this
dataset see][]{romano2007}.

For each $\alpha$-element we are able to reproduce the general trend
of a nearly constant value up to [Fe/H]$\sim$$-$1.0, followed by a
slight decrease down to the value [$\alpha$/Fe]$\sim$$+$0 at
[Fe/H]$\sim$$-$0.6. We also simulate satisfactorily the spread of the
data and the peculiar stars with sub-solar [$\alpha$/Fe] (note in
particular the agreement between observations of [O/Fe] at
[Fe/H]$>-1.6$ and the model prediction).

An alternative explanation for the O and Mg depletion is the formation
of stars from gas polluted by AGB ejecta \citep[c.f.][]{cottrell1981}.
In this contest, self-consistent models of the chemical evolution of
the globular cluster NGC~6752 made by \citet{fenner2004} are not able
to reproduce the O depletion, while Mg turns out to be produced rather
than destroyed; see also \citet{denissenkov2003} 
and \citet{karakas2006b}.

The uncertainties that affect the underlying AGB models undermine the
reliability of the predictions and may leave room for an AGB solution
\citep{ventura2005a, ventura2005b}. The choice of convection treatment
within the AGB models is particularly important.  When convection is
modeled more efficiently, deep Hot Bottom Burning can result
\citep[e.g.]{bloecker1991}, with a strong depletion of oxygen and
magnesium \citep[e.g.][]{ventura2005b,ventura2005c}, while the C+N+O
remains constant to within a factor of two. This could help to explain
the presence of low oxygen abundance stars in the top right panel of
Figure~10.

Since intermediate mass stars (4-8 M$_{\odot}$) evolve over timescales
(30-70 Myr) comparable to that of the peak of SNe~Ia activity (in a
``burst'' environment) we expect that regions inhomogeneously polluted
by SN~Ia should also be enriched by intermediate-mass AGB stars,
complicating the picture. Since AGB stars produces large quantities of
helium \citep[up to Y=0.35, e.g.][]{dantona2005} one might expect
stars polluted by AGB ejecta to be rich in helium, even if the content
hypothesised by several authors to explain the multiple populations in
$\omega$ Cen \citep{piotto2005, lee2005, sollima2007} and in other GC
\citep{dantona2005,piotto2007} require that such stars formed from
``pure'' AGB ejecta. Regardless, \citet{karakas2006} found that only a
maximum helium content of Y$\sim0.26$ was feasible whitin their models,
without violating the observational constraint of near-constant C+N+O.



Finally, $\omega$~Cen shows a clear trend of increasing heavy elements
produced by s-process synthesis (e.g. Rb, Ba, La, and Nd) with
increasing [Fe/H] \citep{norris1995, smith2000}. This is usually interpreted
as a sign of the progressive chemical enrichment from 1.5-5
M$_{\odot}$ AGB stars. The stellar lifetime of stars with mass in
these range \citep[$\sim$ 1 Gyr for a 2 M$_{\odot}$
star][]{schaller1992} are compatible with the SFH inferred by our
study and furthermore the mass reduction in the last $\sim$600 Myr can
aid in the rapid enrichment of these long-lived progenitors'
elements. This kind of s-process enrichment trend is peculiar to
$\omega$~Cen, in comparison with other GC stars.

To be fair, models of the sort described above may only apply to
``traditional'' GCs, and may not be applicable to $\omega$~Cen, but we
felt it prudent to at least discussion the feasibility of this
scenario here. Recall, $\omega$~Cen is the only GC clearly showing a
significant spread in iron abundance (as required by SNe~Ia
enrichment). We are currently developing a new model for canonical GC
evolution based upon similar ingredients (Marcolini et al. in
preparation).

In conclusion, AGB pollution has certainly occurred at some level
within $\omega$~Cen, and cannot be ignored. With this in mind, in the
next section we test the validity of our model by simulating the CMD
of $\omega$~Cen and comparing it with observed CMDs from the
literature.

\begin{figure*}    
\psfig{figure=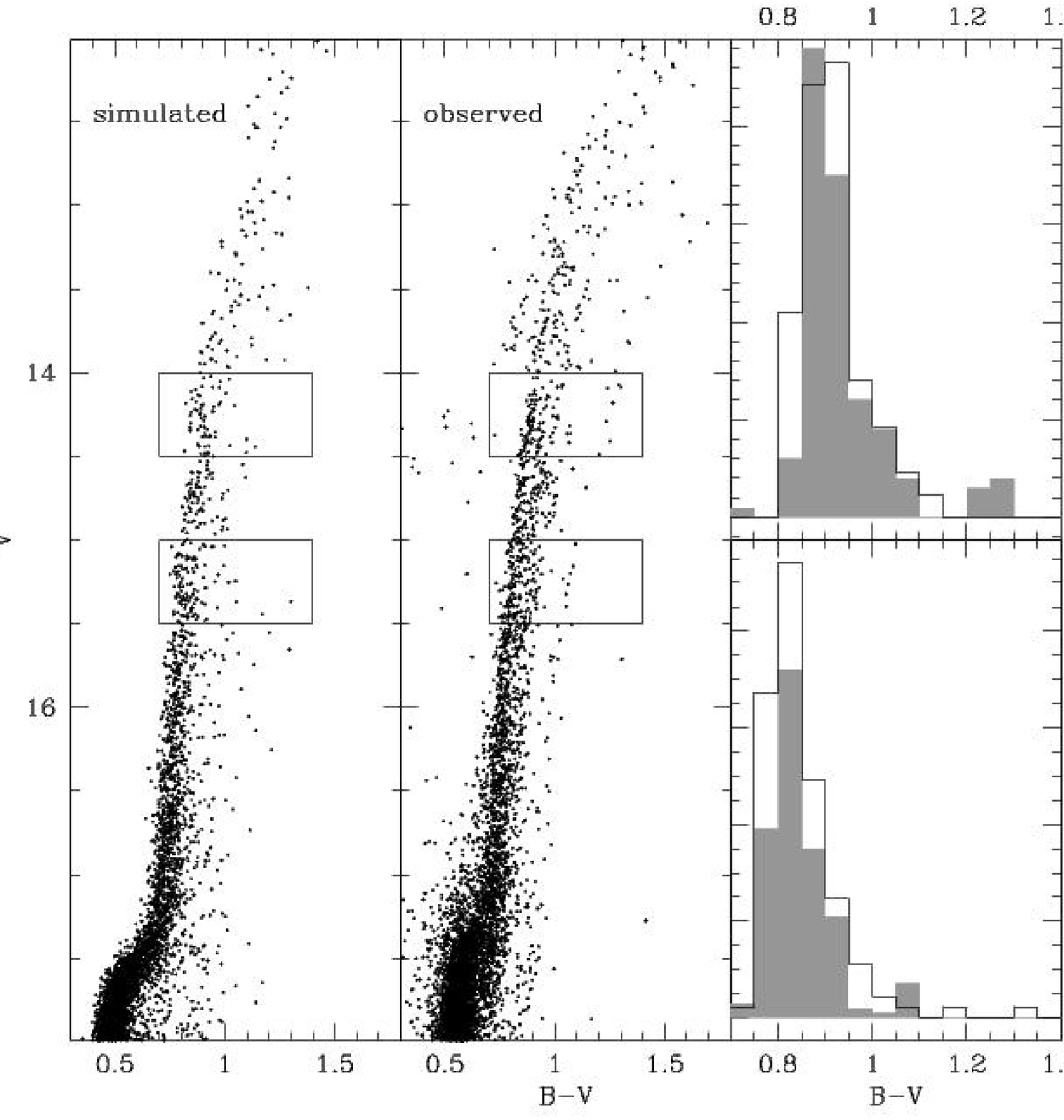,width=0.95\textwidth}    
\caption{Comparison between the simulated (D30LONG model; left panel) and
the observed (from Sollima et al. 2005a; central panel) (V; B-V) CMD
of $\omega$ Cen.  Right panels shows the color histograms between two
different V magnitude intervals (indicated by rectangles in the left
and central panels) for the observed and simulated (grey histograms)
CMD.}
\label{fig:rgb} 
\end{figure*}

\section{Comparing the color-magnitude diagrams}    
\label{sec:comparing_cmd}   

Most of the evidence gathered in the past regarding the metallicity
spread, the presence of multiple populations, and the possible helium
enhancement in $\omega$~Cen have been deduced from the peculiar
morphology of different sequences in the CMD.  It is interesting to
check how our models compare with the most recent and accurate
observational data.  As pointed out in Section~\ref{sec:results}, our
model predicts a wide spread in the distribution of [Fe/H] as well as
[$\alpha$/Fe].  It has been shown that both [Fe/H] and [$\alpha$/Fe]
produce similar effects on the stellar isochrones \citep[][although,
see \citealt{dotter2007} for a more thorough discussion of the impact
of individual elements on stellar population models]{salaris1993,
kim2002, cassisi2004, pietrinferni2006}. In particular, the low
first-ionization potential of Mg, Si and Fe favours the formation of
$H^{-}$ ions that are the main contributors to the mean opacity.
Moreover, the increase in the abundances of C, N, O and Ne produces
significant change in the opacity and in the efficiency of the CNO
nuclear burning that mimic a higher average metallicity $Z$
\citep{renzini1977}.

In Fig.~\ref{fig:rgb} the color distribution of RGB stars at two
different magnitude levels is shown for both the simulated (D30LONG
model) and the observed \citep{sollima2005a} CMD in the V,B-V
plane. The colour distribution of the RGB of $\omega$ Cen is well
reproduced. The most metal-rich anomalous component (RGB-a,
[Fe/H]$\sim$-0.6), well distinguishable in the red part of the CMD of
\citet{sollima2005a}, is less defined in our model (although still
present) because of the large spread in [$\alpha$/Fe] ratios predicted
at this metallicity. However, given the relative simplicity of the
model, we consider the agreement between the simulated and the
observed CMD satisfactory.

\begin{figure}    
\psfig{figure=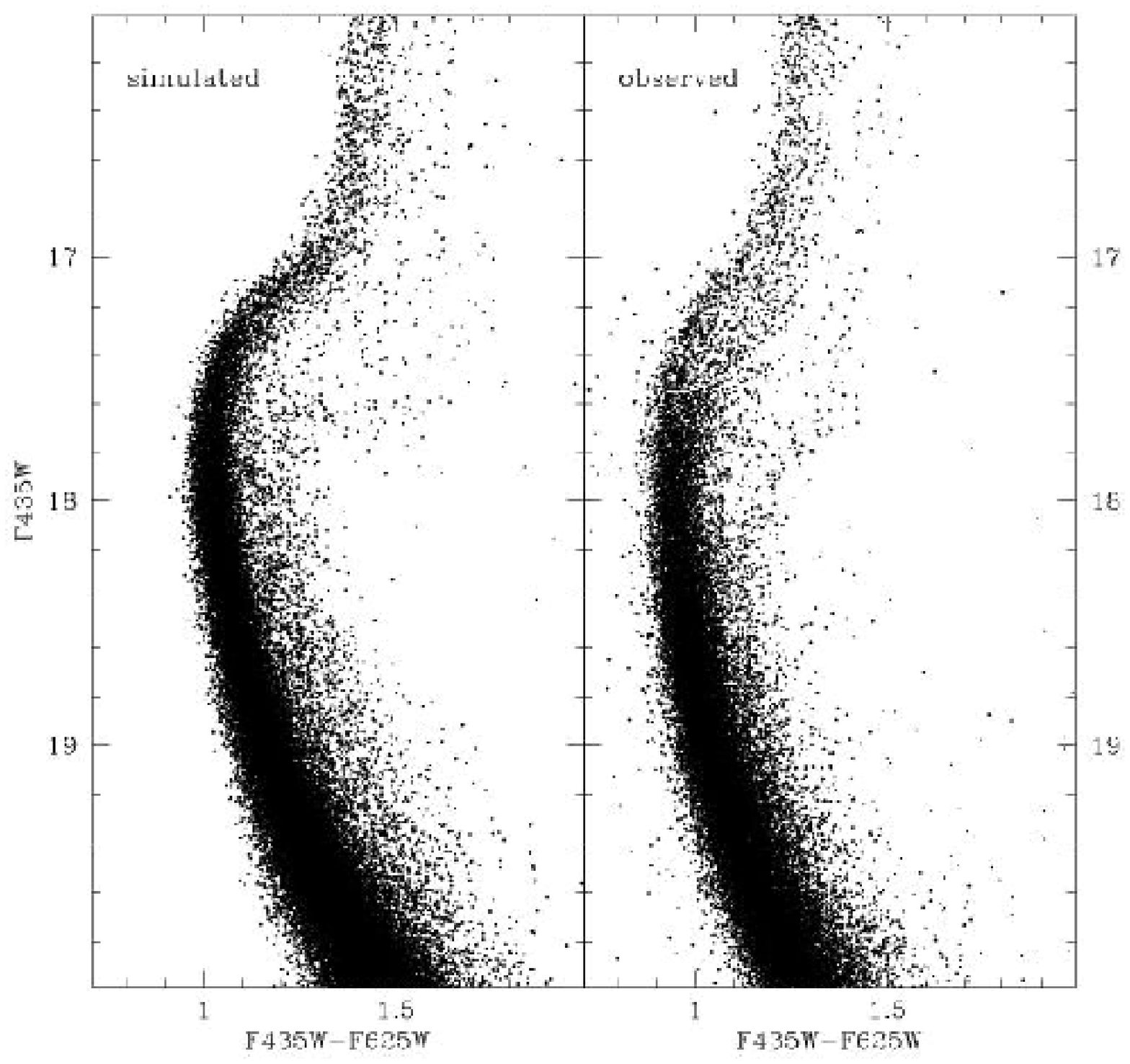,width=0.45\textwidth}
\caption{Comparison between the simulated (D30LONG model; left panel) and 
observed (from Ferraro et al. 2004; right panel) (F625W; F435W-F625W) 
CMD of $\omega$ Cen.}
\label{fig:sgb} 
\end{figure}

A behaviour similar to that observed in Fig.~\ref{fig:rgb} is also
present in Fig.~\ref{fig:sgb}, where our D30LONG model is compared
with the ACS CMD by \citet{ferraro2004} in the sub-giant branch (SGB)
region. As can be noted, the observed magnitude spread of the upper
part of the SGB, as well as the presence of the anomalous SGB (SGB-a),
are accounted for. However, the shape and the slope of the SGB-a are
not well reproduced. This effect has been already discussed by
\citet{ferraro2004} and \citet[][; and references
therein]{sollima2005b} and could be linked to the possible presence of
a significant He overabundance at these metallicities. According to
our model, stars lying along this sequence should have a high iron
content ([Fe/H]$\sim-0.6$), in agreement with spectroscopic
measurements of stars belonging to this sequence \citep{hilker2004,
sollima2005b, stanford2006}.  It is worth recalling that none of our
previous models (DdSph, D90 and D30) reaches such a high metal
content. Indeed, this anomalous metal-rich population appears to form
in the last $\sim$0.6~Gyr of evolution of $\omega$~Cen.

\begin{figure*}    
\psfig{figure=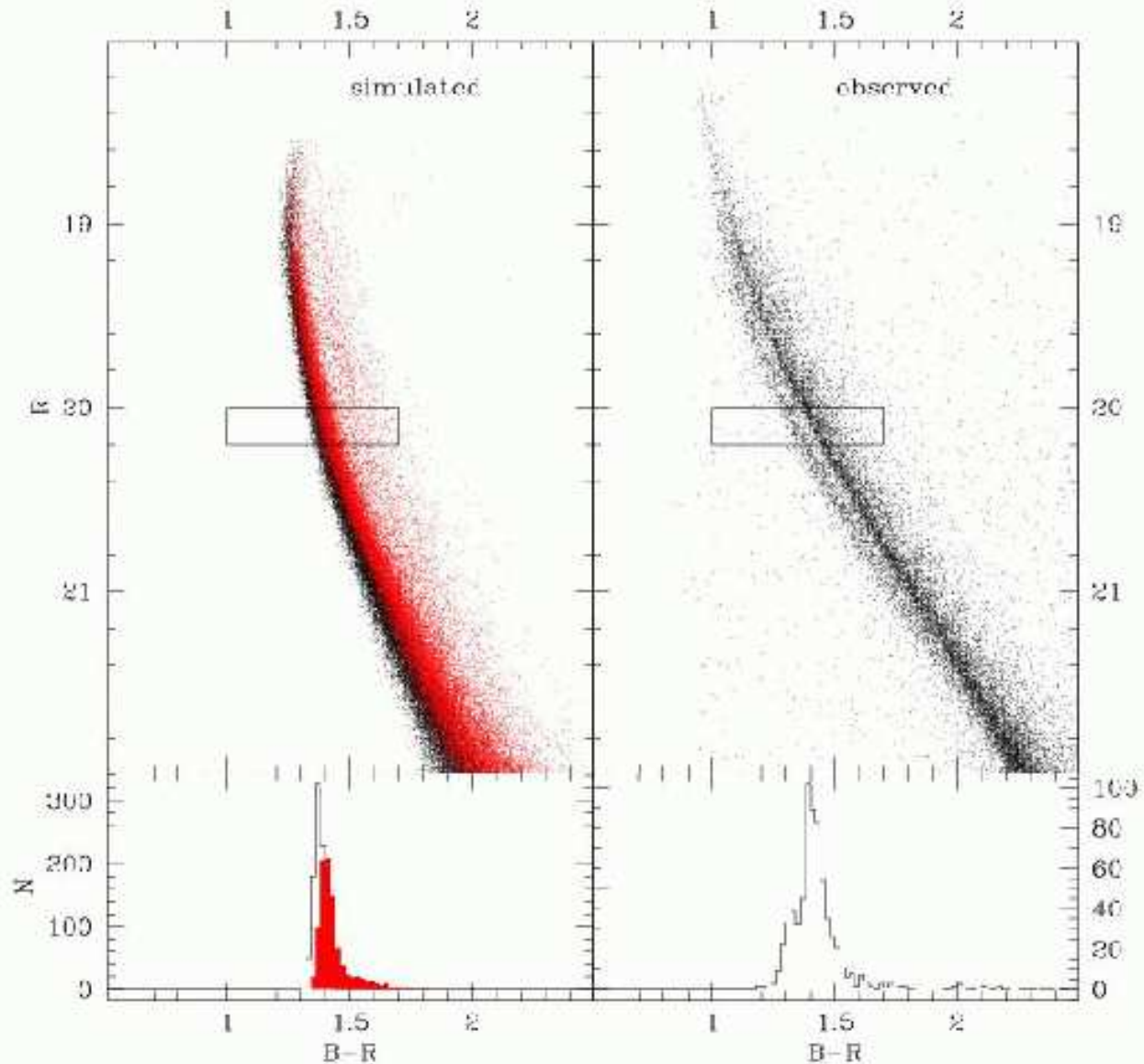,width=0.95\textwidth}    
\caption{Comparison between the simulated (D30LONG model; upper left
  panel) and observed (from Sollima et al. 2007; upper right panel)
  (R; B-R) CMD of $\omega$ Cen. Grey points in the left panel marks
  stars with [Fe/H]$>-1.4$. In the bottom panels the MS colour
  histograms calculated at $20<R<20.2$ are shown.}
\label{fig:ms} 
\end{figure*}

Finally, in Fig.~\ref{fig:ms} we compare our model with the CMD
obtained by \citet{sollima2007} from deep FORS1 observations of the MS
of $\omega$~Cen. Metal-rich stars ([Fe/H]$\gta -1.4$) are marked as
grey points in the synthetic CMD. In the bottom panels, the colour
distributions of Fe-poor ([Fe/H]$<-1.4$) and Fe-rich
([Fe/H]$\gta-1.4$) stars in the simulated CMD and of observed stars
are shown. As can be noted, Fe-rich stars are mostly located in the
same region of the other Fe-poor stars as expected for two population
with similar distributions of Z (see Section~\ref{sec:results}).

As noted in Section~1, spectroscopic analyses indicate that metal-rich
([Fe/H]$\sim-1.3$) MS stars in $\omega$~Cen have been found to lie in
the blue portion of the MS, in contradiction with the canonical
theoretical models \citep{piotto2005}. In order to explain the
anomalous location of these stars in the CMD, a large helium
overabundance has been claimed by several authors \citep{norris2004,
piotto2005, lee2005, sollima2007}. According to our model, the mean
helium abundance remains basically unchanged during the entire process
of self-enrichment at any distance from the cluster centre ($\Delta
Y<0.05$). The same conclusion is drawn from recent analytical
(ie. without associated dynamical evolution of the gas) galactic
chemical evolution models \citep{romano2007}.  Although our model
fails to reproduce the MS morphology of $\omega$~Cen, the lower mean
$\alpha$-element abundance predicted for the Fe-rich group of stars
tends to reduce the color distance between bMS and rMS. According to
the theoretical models of \citet{straniero1997}, in this scenario the
required amount of helium overabundance to shift the metal-rich
([Fe/H]$\sim-1.3$) MS on the blue side of the dominant Fe-poor cluster
MS is reduced by a factor of 40\% ($\Delta Y\sim 0.09$).

\begin{figure}    
\psfig{figure=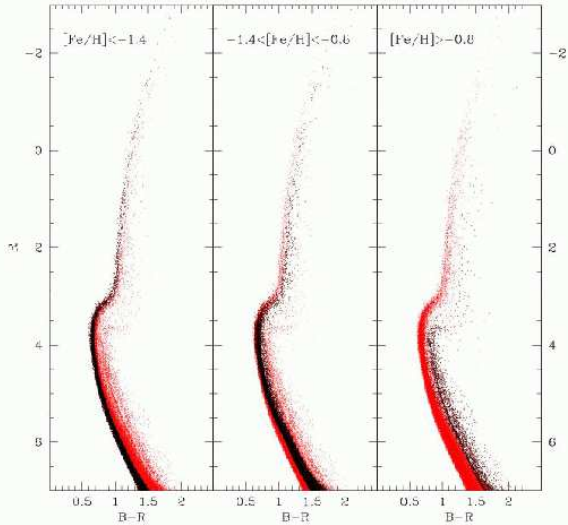,width=0.45\textwidth}    
\caption{Simulated (R; B-R) CMD of $\omega$ Cen (D30LONG model). In the
different panels stars with [Fe/H]$<$-1.4 (left panel),
-$1.4<$[Fe/H]$<$-0.8 (central panel) and [Fe/H]$>$-0.8 (right panel) are
marked as black points.}
\label{fig:metcmd} 
\end{figure}

Finally, in Fig.~\ref{fig:metcmd} we plot our full (R,B-R) CMD.  The
location in the CMD of stars in three different ranges of [Fe/H] are
shown. Note that, although the three groups populate preferentially
different regions of the CMD, they partially overlap. As discussed in
Section~\ref{sec:results}, this is due to the fact that the [Fe/H] is
not stictly linearly proportional to the metal mass fraction, and
stars with the same $Z$ can have different [Fe/H] content. In this
picture our model could explain the peculiar position of the outliers
observed by \citet{villanova2007} in their spectroscopic analysis of
sub-giant stars. These stars have been indeed found to have a
[Fe/H]$>$-1.0 although their position in the CMD would suggest a
significant smaller [Fe/H]. The comparison with our model suggests
that the location of these stars in the CMD could be explained if
these stars show a significant [$\alpha$/Fe] depletion.

While a detailed comparison of every aspect of the CMD is beyond the
scope of the paper, we have demonstrated that the possibility that
$\omega$~Cen had its origin in a single progression of star formation
and metal enrichment cannot yet be entirely dismissed (cf. Villanova
et~al. 2007).  Our proposed scenario is not meant to be interpreted as
the only solution to the problem and to recover all the aspects of the
CMD morphology, but neither should it be ignored when interpreting the
MS, SGB and RGB morphologies (it may be that a required solution
involving our mechanism alongside that of enhanced helium may be
required).

\section{Discussion and Conclusions}
\label{sec:discussion}

We studied the chemical evolution of the peculiar stellar system
$\omega$~Cen under the assumption that it represents the nuclear
region of a dSph deprived of most of its mass by its interaction with
the Milky Way. We based our model on hydrodynamical simulations by
Marcolini et al. (2006) describing the evolution of an isolated dSph
galaxy. We focused our analysis on the central region of the system,
comparing its chemical properties with the observational data
available in literature.

We paid particular attention to the different role played by SNe~II
and SNe~Ia in the chemical enrichment of the gas. While the SNe~II
pollute the ISM rather uniformly, the SN~Ia ejecta may remain
temporarily confined inside relatively small pockets of gas before
being mixed with the ISM. The stars forming in such pockets have lower
[$\alpha$/Fe] ratios than the stars forming elsewhere which show a
significant $\alpha$-enhancement ($\langle[\alpha$/Fe]$\rangle$
$\ge$0.35). $\alpha$-depleted stars represent only $\sim$7\% of the
entire dSph stellar population, but their fraction rises up to
$\sim$30\% in the central region. This difference depends on the
hydrodynamical evolution, as well as on the higher rate of SNe Ia in
the central regions of the system. As a consequence, the chemical
properties of the stars forming in the nuclear region of a dSph differ
substantially from those of the main dSph population, and can resemble
those observed in $\omega$ Cen.

The $Z$ and [Fe/H] distributions of $\omega$ Cen have been reproduced
assuming that 85\% of the stars were born within 1~Gyr (in a sequence
of equal-intensity starbursts) and the remaining fraction formed
during the subsequent 0.6~Gyr at a lower rate (owing to the
interaction with the Galaxy). The stars created in this last period
show a rapid increase of their [Fe/H] because of the rapid reduction
of the gas mass content as a consequence of the interaction between
the dSph and the Milky Way.

The first constraint against wich our model has been tested is that
of the [Fe/H] and [Ca/H] distributions of the stars observed in
$\omega$ Cen \citep{norris1996, sollima2005a}.  The inhomogeneous
mixing of the metals ejected by SNe~Ia naturally explains the observed
multimodal distributions, as well as their radial profile.

From accurate CMD simulations, we compared the morphology of the
evolutionary sequences with that predicted by our model.  
Our model is successful in reproducing the overall spread in the 
RGB and SGB morphology, but fails to mimic the SGB-a morphology.
It also fails to reproduce the bimodal MS observed in this
cluster.  However, due to the relationship between the [$\alpha$/Fe]
and [Fe/H] ratios in the SNe~Ia pockets, the discrepancy between the
predicted and observed location of the bMS component can be
ameliorated by requiring a reduced level to the helium overabundance
($\Delta~Y\sim 0.09$) usually invoked \citep{norris2004, piotto2005,
sollima2007}.

A further test able to validate the inhomogeneous pollution expected
by our model should be the observation in $\omega$~Cen of some stars
with sub-solar $[\alpha$/Fe] ratios (at [Fe/H]$ \ge-1.5$). Note that
stars with such a peculiar abundance patterns have already been
observed in the centre of some dSphs like Sagittarius
\citep{bonifacio2004, monaco2005, mcwilliam2005, sbordone2007}, Draco
\citep{shetrone2001} ,Fornax and Sculptor \citep{venn2005,tolstoy2006}.

We note that in order to realise our synthetic CMDs, we utilised
evolutionary tracks \citep{cassisi2004, pietrinferni2006} calculated
for two different $\alpha$-enhancement levels ([$\alpha$/Fe]=$+$0.0
and [$\alpha$/Fe]=$+$0.4). We then extrapolated down to the ratio
[$\alpha$/Fe]=$-$0.4 (for each $Z$ value) in order to take into
account the entire [$\alpha$/Fe] range covered by the stars in our
model. Unfortunatelty, most of these stars do not have solar-scaled
abundances, as assumed in the utilised isochrones. Tracks taking into
account the relation among $Z$, [Fe/H] and [$\alpha$/Fe] given by
equation (1) would be desirable to obtain more realistic
tests. Moreover, further models taking into account the dynamical
interaction of the parent dwarf galaxy with the Milky Way and the
effect of AGB pollution will be required.

\section*{Acknowledgments}

We are grateful to the anonymous referee for his/her helpful
suggestions which improved the presentation of the paper.  We
acknowledge financial support from National Institute for Astrophysics
(INAF). The simulations were run at the CINECA Supercomputing Centre
with CPU time assigned thanks to INAF-CINECA grant. We thank Santino
Cassisi for the helpful discussion on the $\alpha$-enhanced
evolutionary tracks. We also thank Lars Freyhammer, Leticia Carigi and
Giuseppina Battaglia for reading the manuscript, as well as for
providing useful feedback and suggestions.
    

\bibliographystyle{mn2e}    
\bibliography{omega_refs}    

\begin{thebibliography}{85}
\expandafter\ifx\csname natexlab\endcsname\relax\def\natexlab#1{#1}\fi

\bibitem[{{Anderson}(2002)}]{anderson2002}
{Anderson} J., 2002, in Astronomical Society of the Pacific Conference Series,
  Vol. 265, Omega Centauri, A Unique Window into Astrophysics, {van Leeuwen}
  F., {Hughes} J.~D., {Piotto} G., eds., p.~87

\bibitem[{{Babusiaux} {et~al.}(2005){Babusiaux}, {Gilmore}, \&
  {Irwin}}]{babusiaux2005}
{Babusiaux} C., {Gilmore} G., {Irwin} M., 2005, \mnras, 359, 985

\bibitem[{{Battaglia} {et~al.}(2006){Battaglia}, {Tolstoy}, {Helmi}, {Irwin},
  {Letarte}, {Jablonka}, {Hill}, {Venn}, {Shetrone}, {Arimoto}, {Primas},
  {Kaufer}, {Francois}, {Szeifert}, {Abel}, \& {Sadakane}}]{battaglia2006}
{Battaglia} G., {Tolstoy} E., {Helmi} A., {Irwin} M.~J., {Letarte} B.,
  {Jablonka} P., {Hill} V., {Venn} K.~A., {Shetrone} M.~D., {Arimoto} N.,
  {Primas} F., {Kaufer} A., {Francois} P., {Szeifert} T., {Abel} T., {Sadakane}
  K., 2006, \aap, 459, 423

\bibitem[{{Bedin} {et~al.}(2004){Bedin}, {Piotto}, {Anderson}, {Cassisi},
  {King}, {Momany}, \& {Carraro}}]{bedin2004}
{Bedin} L.~R., {Piotto} G., {Anderson} J., {Cassisi} S., {King} I.~R., {Momany}
  Y., {Carraro} G., 2004, \apjl, 605, L125

\bibitem[{{Bekki} \& {Freeman}(2003)}]{bekki2003}
{Bekki} K., {Freeman} K.~C., 2003, \mnras, 346, L11

\bibitem[{{Bekki} \& {Norris}(2006)}]{bekki2006}
{Bekki} K., {Norris} J.~E., 2006, \apjl, 637, L109

\bibitem[{{Bellazzini} {et~al.}(2002){Bellazzini}, {Ferraro}, {Origlia},
  {Pancino}, {Monaco}, \& {Oliva}}]{bellazzini2002}
{Bellazzini} M., {Ferraro} F.~R., {Origlia} L., {Pancino} E., {Monaco} L.,
  {Oliva} E., 2002, \aj, 124, 3222

\bibitem[{{Bellazzini} {et~al.}(2004){Bellazzini}, {Ferraro}, {Sollima},
  {Pancino}, \& {Origlia}}]{bellazzini2004}
{Bellazzini} M., {Ferraro} F.~R., {Sollima} A., {Pancino} E., {Origlia} L.,
  2004, \aap, 424, 199

\bibitem[{{Bloecker} \& {Schoenberner}(1991)}]{bloecker1991}
{Bloecker} T., {Schoenberner} D., 1991, \aap, 244, L43

\bibitem[{{Bonifacio} {et~al.}(2004){Bonifacio}, {Sbordone}, {Marconi},
  {Pasquini}, \& {Hill}}]{bonifacio2004}
{Bonifacio} P., {Sbordone} L., {Marconi} G., {Pasquini} L., {Hill} V., 2004,
  \aap, 414, 503

\bibitem[{{Bosler} {et~al.}(2007){Bosler}, {Smecker-Hane}, \&
  {Stetson}}]{bosler2007}
{Bosler} T.~L., {Smecker-Hane} T.~A., {Stetson} P.~B., 2007, \mnras, 378, 318

\bibitem[{{Brown} \& {Wallerstein}(1993)}]{brown1993}
{Brown} J.~A., {Wallerstein} G., 1993, \aj, 106, 133

\bibitem[{{Cannon} \& {Stobie}(1973)}]{cannon1973}
{Cannon} R.~D., {Stobie} R.~S., 1973, \mnras, 162, 207

\bibitem[{{Carraro} \& {Lia}(2000)}]{carraro2000}
{Carraro} G., {Lia} C., 2000, \aap, 357, 977

\bibitem[{{Cassisi} {et~al.}(2004){Cassisi}, {Salaris}, {Castelli}, \&
  {Pietrinferni}}]{cassisi2004}
{Cassisi} S., {Salaris} M., {Castelli} F., {Pietrinferni} A., 2004, \apj, 616,
  498

\bibitem[{{Cottrell} \& {Da Costa}(1981)}]{cottrell1981}
{Cottrell} P.~L., {Da Costa} G.~S., 1981, \apjl, 245, L79

\bibitem[{{D'Antona} {et~al.}(2005){D'Antona}, {Bellazzini}, {Caloi}, {Pecci},
  {Galleti}, \& {Rood}}]{dantona2005}
{D'Antona} F., {Bellazzini} M., {Caloi} V., {Pecci} F.~F., {Galleti} S., {Rood}
  R.~T., 2005, \apj, 631, 868

\bibitem[{{Del Principe} {et~al.}(2006){Del Principe}, {Piersimoni}, {Storm},
  {Caputo}, {Bono}, {Stetson}, {Castellani}, {Buonanno}, {Calamida}, {Corsi},
  {Dall'Ora}, {Ferraro}, {Freyhammer}, {Iannicola}, {Monelli}, {Nonino},
  {Pulone}, \& {Ripepi}}]{delprincipe2006}
{Del Principe} M., {Piersimoni} A.~M., {Storm} J., {Caputo} F., {Bono} G.,
  {Stetson} P.~B., {Castellani} M., {Buonanno} R., {Calamida} A., {Corsi}
  C.~E., {Dall'Ora} M., {Ferraro} I., {Freyhammer} L.~M., {Iannicola} G.,
  {Monelli} M., {Nonino} M., {Pulone} L., {Ripepi} V., 2006, \apj, 652, 362

\bibitem[{{Denissenkov} \& {Herwig}(2003)}]{denissenkov2003}
{Denissenkov} P.~A., {Herwig} F., 2003, \apjl, 590, L99

\bibitem[{{Dinescu} {et~al.}(1999){Dinescu}, {van Altena}, {Girard}, \&
  {L{\'o}pez}}]{dinescu1999}
{Dinescu} D.~I., {van Altena} W.~F., {Girard} T.~M., {L{\'o}pez} C.~E., 1999,
  \aj, 117, 277

\bibitem[{{Dotter} {et~al.}(2007){Dotter}, {Chaboyer}, {Ferguson}, {Lee},
  {Worthey}, {Baron}, \& {Jevremovic}}]{dotter2007}
{Dotter} A., {Chaboyer} B., {Ferguson} J.~W., {Lee} H.~., {Worthey} G., {Baron}
  E., {Jevremovic} D., 2007, astro-ph/07060808

\bibitem[{{Edvardsson} {et~al.}(1993){Edvardsson}, {Andersen}, {Gustafsson},
  {Lambert}, {Nissen}, \& {Tomkin}}]{edvardsson1993}
{Edvardsson} B., {Andersen} J., {Gustafsson} B., {Lambert} D.~L., {Nissen}
  P.~E., {Tomkin} J., 1993, \aaps, 102, 603

\bibitem[{{Fenner} {et~al.}(2004){Fenner}, {Campbell}, {Karakas}, {Lattanzio},
  \& {Gibson}}]{fenner2004}
{Fenner} Y., {Campbell} S., {Karakas} A.~I., {Lattanzio} J.~C., {Gibson} B.~K.,
  2004, \mnras, 353, 789

\bibitem[{{Ferraro} {et~al.}(2004){Ferraro}, {Sollima}, {Pancino},
  {Bellazzini}, {Straniero}, {Origlia}, \& {Cool}}]{ferraro2004}
{Ferraro} F.~R., {Sollima} A., {Pancino} E., {Bellazzini} M., {Straniero} O.,
  {Origlia} L., {Cool} A.~M., 2004, \apjl, 603, L81

\bibitem[{{Ferraro} {et~al.}(2006){Ferraro}, {Sollima}, {Rood}, {Origlia},
  {Pancino}, \& {Bellazzini}}]{ferraro2006}
{Ferraro} F.~R., {Sollima} A., {Rood} R.~T., {Origlia} L., {Pancino} E.,
  {Bellazzini} M., 2006, \apj, 638, 433

\bibitem[{{Francois} {et~al.}(1988){Francois}, {Spite}, \&
  {Spite}}]{francois1988}
{Francois} P., {Spite} M., {Spite} F., 1988, \aap, 191, 267

\bibitem[{{Gnedin} {et~al.}(2002){Gnedin}, {Zhao}, {Pringle}, {Fall}, {Livio},
  \& {Meylan}}]{gnedin2002}
{Gnedin} O.~Y., {Zhao} H., {Pringle} J.~E., {Fall} S.~M., {Livio} M., {Meylan}
  G., 2002, \apjl, 568, L23

\bibitem[{{Goswami} \& {Prantzos}(2000)}]{Goswami2000}
{Goswami} A., {Prantzos} N., 2000, \aap, 359, 191

\bibitem[{{Grevesse} \& {Sauval}(1998)}]{grevesse1998}
{Grevesse} N., {Sauval} A.~J., 1998, Space Science Reviews, 85, 161

\bibitem[{{Hilker} {et~al.}(2004){Hilker}, {Kayser}, {Richtler}, \&
  {Willemsen}}]{hilker2004}
{Hilker} M., {Kayser} A., {Richtler} T., {Willemsen} P., 2004, \aap, 422, L9

\bibitem[{{Hughes} \& {Wallerstein}(2000)}]{hughes2000}
{Hughes} J., {Wallerstein} G., 2000, \aj, 119, 1225

\bibitem[{{Iwamoto} {et~al.}(1999){Iwamoto}, {Brachwitz}, {Nomoto},
  {Kishimoto}, {Umeda}, {Hix}, \& {Thielemann}}]{iwamoto1999}
{Iwamoto} K., {Brachwitz} F., {Nomoto} K., {Kishimoto} N., {Umeda} H., {Hix}
  W.~R., {Thielemann} F.-K., 1999, \apjs, 125, 439

\bibitem[{{Karakas} {et~al.}(2006{\natexlab{a}}){Karakas}, {Fenner}, {Sills},
  {Campbell}, \& {Lattanzio}}]{karakas2006}
{Karakas} A.~I., {Fenner} Y., {Sills} A., {Campbell} S.~W., {Lattanzio} J.~C.,
  2006{\natexlab{a}}, \apj, 652, 1240

\bibitem[{{Karakas} {et~al.}(2006{\natexlab{b}}){Karakas}, {Lugaro}, {Ugalde},
  {Wiescher}, \& {G{\"o}rres}}]{karakas2006b}
{Karakas} A.~I., {Lugaro} M., {Ugalde} C., {Wiescher} M., {G{\"o}rres} J.,
  2006{\natexlab{b}}, New Astronomy Review, 50, 500

\bibitem[{{Kim} {et~al.}(2002){Kim}, {Demarque}, {Yi}, \&
  {Alexander}}]{kim2002}
{Kim} Y.-C., {Demarque} P., {Yi} S.~K., {Alexander} D.~R., 2002, \apjs, 143,
  499

\bibitem[{{Koch} {et~al.}(2006){Koch}, {Grebel}, {Wyse}, {Kleyna}, {Wilkinson},
  {Harbeck}, {Gilmore}, \& {Evans}}]{koch2006}
{Koch} A., {Grebel} E.~K., {Wyse} R.~F.~G., {Kleyna} J.~T., {Wilkinson} M.~I.,
  {Harbeck} D.~R., {Gilmore} G.~F., {Evans} N.~W., 2006, \aj, 131, 895

\bibitem[{{Lee} {et~al.}(2005){Lee}, {Joo}, {Han}, {Chung}, {Ree}, {Sohn},
  {Kim}, {Yoon}, {Yi}, \& {Demarque}}]{lee2005}
{Lee} Y.-W., {Joo} S.-J., {Han} S.-I., {Chung} C., {Ree} C.~H., {Sohn} Y.-J.,
  {Kim} Y.-C., {Yoon} S.-J., {Yi} S.~K., {Demarque} P., 2005, \apjl, 621, L57

\bibitem[{{Limongi} {et~al.}(2000){Limongi}, {Straniero}, \&
  {Chieffi}}]{limongi2000}
{Limongi} M., {Straniero} O., {Chieffi} A., 2000, \apjs, 129, 625

\bibitem[{{Lub}(2002)}]{lub2002}
{Lub} J., 2002, in ASP Conf. Ser. 265: Omega Centauri, A Unique Window into
  Astrophysics, {van Leeuwen} F., {Hughes} J.~D., {Piotto} G., eds., p.~95

\bibitem[{{Majewski} {et~al.}(2000){Majewski}, {Patterson}, {Dinescu},
  {Johnson}, {Ostheimer}, {Kunkel}, \& {Palma}}]{majewski2000}
{Majewski} S.~R., {Patterson} R.~J., {Dinescu} D.~I., {Johnson} W.~Y.,
  {Ostheimer} J.~C., {Kunkel} W.~E., {Palma} C., 2000, in Liege International
  Astrophysical Colloquia, {Noels} A., {Magain} P., {Caro} D., {Jehin} E.,
  {Parmentier} G., {Thoul} A.~A., eds., p. 619

\bibitem[{{Marcolini} {et~al.}(2003){Marcolini}, {Brighenti}, \&
  {D'Ercole}}]{marcolini2003}
{Marcolini} A., {Brighenti} F., {D'Ercole} A., 2003, \mnras, 345, 1329

\bibitem[{{Marcolini} {et~al.}(2006){Marcolini}, {D'Ercole}, {Brighenti}, \&
  {Recchi}}]{marcolini2006}
{Marcolini} A., {D'Ercole} A., {Brighenti} F., {Recchi} S., 2006, \mnras, 371,
  643

\bibitem[{{Mateo}(1998)}]{mateo1998}
{Mateo} M.~L., 1998, \araa, 36, 435

\bibitem[{{Mayer} {et~al.}(2006){Mayer}, {Mastropietro}, {Wadsley}, {Stadel},
  \& {Moore}}]{mayer2006}
{Mayer} L., {Mastropietro} C., {Wadsley} J., {Stadel} J., {Moore} B., 2006,
  \mnras, 369, 1021

\bibitem[{{McWilliam} \& {Smecker-Hane}(2005)}]{mcwilliam2005}
{McWilliam} A., {Smecker-Hane} T.~A., 2005, in Astronomical Society of the
  Pacific Conference Series, Vol. 336, Cosmic Abundances as Records of Stellar
  Evolution and Nucleosynthesis, {Barnes} III T.~G., {Bash} F.~N., eds., p. 221

\bibitem[{{Meylan} {et~al.}(1995){Meylan}, {Mayor}, {Duquennoy}, \&
  {Dubath}}]{meylan1995}
{Meylan} G., {Mayor} M., {Duquennoy} A., {Dubath} P., 1995, \aap, 303, 761

\bibitem[{{Monaco} {et~al.}(2005){Monaco}, {Bellazzini}, {Bonifacio},
  {Ferraro}, {Marconi}, {Pancino}, {Sbordone}, \& {Zaggia}}]{monaco2005}
{Monaco} L., {Bellazzini} M., {Bonifacio} P., {Ferraro} F.~R., {Marconi} G.,
  {Pancino} E., {Sbordone} L., {Zaggia} S., 2005, \aap, 441, 141

\bibitem[{{Norris}(2004)}]{norris2004}
{Norris} J.~E., 2004, \apjl, 612, L25

\bibitem[{{Norris} \& {Da Costa}(1995)}]{norris1995}
{Norris} J.~E., {Da Costa} G.~S., 1995, \apj, 447, 680

\bibitem[{{Norris} {et~al.}(1997){Norris}, {Freeman}, {Mayor}, \&
  {Seitzer}}]{norris1997}
{Norris} J.~E., {Freeman} K.~C., {Mayor} M., {Seitzer} P., 1997, \apjl, 487,
  L187

\bibitem[{{Norris} {et~al.}(1996){Norris}, {Freeman}, \&
  {Mighell}}]{norris1996}
{Norris} J.~E., {Freeman} K.~C., {Mighell} K.~J., 1996, \apj, 462, 241

\bibitem[{{Origlia} {et~al.}(2003){Origlia}, {Ferraro}, {Bellazzini}, \&
  {Pancino}}]{origlia2003}
{Origlia} L., {Ferraro} F.~R., {Bellazzini} M., {Pancino} E., 2003, \apj, 591,
  916

\bibitem[{{Pancino} {et~al.}(2000){Pancino}, {Ferraro}, {Bellazzini}, {Piotto},
  \& {Zoccali}}]{pancino2000}
{Pancino} E., {Ferraro} F.~R., {Bellazzini} M., {Piotto} G., {Zoccali} M.,
  2000, \apjl, 534, L83

\bibitem[{{Pancino} {et~al.}(2002){Pancino}, {Pasquini}, {Hill}, {Ferraro}, \&
  {Bellazzini}}]{pancino2002}
{Pancino} E., {Pasquini} L., {Hill} V., {Ferraro} F.~R., {Bellazzini} M., 2002,
  \apjl, 568, L101

\bibitem[{{Pietrinferni} {et~al.}(2006){Pietrinferni}, {Cassisi}, {Salaris}, \&
  {Castelli}}]{pietrinferni2006}
{Pietrinferni} A., {Cassisi} S., {Salaris} M., {Castelli} F., 2006, \apj, 642,
  797

\bibitem[{{Piotto} {et~al.}(2007){Piotto}, {Bedin}, {Anderson}, {King},
  {Cassisi}, {Milone}, {Villanova}, {Pietrinferni}, \& {Renzini}}]{piotto2007}
{Piotto} G., {Bedin} L.~R., {Anderson} J., {King} I.~R., {Cassisi} S., {Milone}
  A.~P., {Villanova} S., {Pietrinferni} A., {Renzini} A., 2007, \apjl, 661, L53

\bibitem[{{Piotto} {et~al.}(2005){Piotto}, {Villanova}, {Bedin}, {Gratton},
  {Cassisi}, {Momany}, {Recio-Blanco}, {Lucatello}, {Anderson}, {King},
  {Pietrinferni}, \& {Carraro}}]{piotto2005}
{Piotto} G., {Villanova} S., {Bedin} L.~R., {Gratton} R., {Cassisi} S.,
  {Momany} Y., {Recio-Blanco} A., {Lucatello} S., {Anderson} J., {King} I.~R.,
  {Pietrinferni} A., {Carraro} G., 2005, \apj, 621, 777

\bibitem[{{Renzini}(1977)}]{renzini1977}
{Renzini} A., 1977, in Saas-Fee Advanced Course 7: Advanced Stages in Stellar
  Evolution, {Bouvier} P., {Maeder} A., eds., p. 151

\bibitem[{{Rey} {et~al.}(2004){Rey}, {Lee}, {Ree}, {Joo}, {Sohn}, \&
  {Walker}}]{rey2004}
{Rey} S.-C., {Lee} Y.-W., {Ree} C.~H., {Joo} J.-M., {Sohn} Y.-J., {Walker}
  A.~R., 2004, \aj, 127, 958

\bibitem[{{Romano} {et~al.}(2007){Romano}, {Matteucci}, {Tosi}, {Pancino},
  {Bellazzini}, {Ferraro}, {Limongi}, \& {Sollima}}]{romano2007}
{Romano} D., {Matteucci} F., {Tosi} M., {Pancino} E., {Bellazzini} M.,
  {Ferraro} F.~R., {Limongi} M., {Sollima} A., 2007, \mnras, 69

\bibitem[{{Salaris} {et~al.}(1993){Salaris}, {Chieffi}, \&
  {Straniero}}]{salaris1993}
{Salaris} M., {Chieffi} A., {Straniero} O., 1993, \apj, 414, 580

\bibitem[{{Savage} \& {Mathis}(1979)}]{savage1979}
{Savage} B.~D., {Mathis} J.~S., 1979, \araa, 17, 73

\bibitem[{{Sbordone} {et~al.}(2007){Sbordone}, {Bonifacio}, {Buonanno},
  {Marconi}, {Monaco}, \& {Zaggia}}]{sbordone2007}
{Sbordone} L., {Bonifacio} P., {Buonanno} R., {Marconi} G., {Monaco} L.,
  {Zaggia} S., 2007, \aap, 465, 815

\bibitem[{{Schaller} {et~al.}(1992){Schaller}, {Schaerer}, {Meynet}, \&
  {Maeder}}]{schaller1992}
{Schaller} G., {Schaerer} D., {Meynet} G., {Maeder} A., 1992, \aaps, 96, 269

\bibitem[{{Shetrone} {et~al.}(2001){Shetrone}, {C{\^o}t{\'e}}, \&
  {Sargent}}]{shetrone2001}
{Shetrone} M.~D., {C{\^o}t{\'e}} P., {Sargent} W.~L.~W., 2001, \apj, 548, 592

\bibitem[{{Sirianni} {et~al.}(2005){Sirianni}, {Jee}, {Ben{\'{\i}}tez},
  {Blakeslee}, {Martel}, {Meurer}, {Clampin}, {De Marchi}, {Ford}, {Gilliland},
  {Hartig}, {Illingworth}, {Mack}, \& {McCann}}]{sirianni2005}
{Sirianni} M., {Jee} M.~J., {Ben{\'{\i}}tez} N., {Blakeslee} J.~P., {Martel}
  A.~R., {Meurer} G., {Clampin} M., {De Marchi} G., {Ford} H.~C., {Gilliland}
  R., {Hartig} G.~F., {Illingworth} G.~D., {Mack} J., {McCann} W.~J., 2005,
  \pasp, 117, 1049

\bibitem[{{Smith} {et~al.}(1995){Smith}, {Cunha}, \& {Lambert}}]{smith1995}
{Smith} V.~V., {Cunha} K., {Lambert} D.~L., 1995, \aj, 110, 2827

\bibitem[{{Smith} {et~al.}(2000){Smith}, {Suntzeff}, {Cunha}, {Gallino},
  {Busso}, {Lambert}, \& {Straniero}}]{smith2000}
{Smith} V.~V., {Suntzeff} N.~B., {Cunha} K., {Gallino} R., {Busso} M.,
  {Lambert} D.~L., {Straniero} O., 2000, \aj, 119, 1239

\bibitem[{{Sollima} {et~al.}(2007){Sollima}, {Ferraro}, {Bellazzini},
  {Origlia}, {Straniero}, \& {Pancino}}]{sollima2007}
{Sollima} A., {Ferraro} F.~R., {Bellazzini} M., {Origlia} L., {Straniero} O.,
  {Pancino} E., 2007, \apj, 654, 915

\bibitem[{{Sollima} {et~al.}(2005{\natexlab{a}}){Sollima}, {Ferraro},
  {Pancino}, \& {Bellazzini}}]{sollima2005a}
{Sollima} A., {Ferraro} F.~R., {Pancino} E., {Bellazzini} M.,
  2005{\natexlab{a}}, \mnras, 357, 265

\bibitem[{{Sollima} {et~al.}(2005{\natexlab{b}}){Sollima}, {Pancino},
  {Ferraro}, {Bellazzini}, {Straniero}, \& {Pasquini}}]{sollima2005b}
{Sollima} A., {Pancino} E., {Ferraro} F.~R., {Bellazzini} M., {Straniero} O.,
  {Pasquini} L., 2005{\natexlab{b}}, \apj, 634, 332

\bibitem[{{Stanford} {et~al.}(2006){Stanford}, {Da Costa}, {Norris}, \&
  {Cannon}}]{stanford2006}
{Stanford} L.~M., {Da Costa} G.~S., {Norris} J.~E., {Cannon} R.~D., 2006, \apj,
  647, 1075

\bibitem[{{Straniero} {et~al.}(1997){Straniero}, {Chieffi}, \&
  {Limongi}}]{straniero1997}
{Straniero} O., {Chieffi} A., {Limongi} M., 1997, \apj, 490, 425

\bibitem[{{Suntzeff} \& {Kraft}(1996)}]{suntzeff1996}
{Suntzeff} N.~B., {Kraft} R.~P., 1996, \aj, 111, 1913

\bibitem[{{Tolstoy} {et~al.}(2006){Tolstoy}, {Hill}, {Irwin}, {Helmi},
  {Battaglia}, {Letarte}, {Venn}, {Jablonka}, {Shetrone}, {Arimoto}, {Abel},
  {Primas}, {Kaufer}, {Szeifert}, {Francois}, \& {Sadakane}}]{tolstoy2006}
{Tolstoy} E., {Hill} V., {Irwin} M., {Helmi} A., {Battaglia} G., {Letarte} B.,
  {Venn} K., {Jablonka} P., {Shetrone} M., {Arimoto} N., {Abel} T., {Primas}
  F., {Kaufer} A., {Szeifert} T., {Francois} P., {Sadakane} K., 2006, The
  Messenger, 123, 33

\bibitem[{{Tsuchiya} {et~al.}(2004){Tsuchiya}, {Korchagin}, \&
  {Dinescu}}]{tsuchiya2004}
{Tsuchiya} T., {Korchagin} V.~I., {Dinescu} D.~I., 2004, \mnras, 350, 1141

\bibitem[{{van de Ven} {et~al.}(2006){van de Ven}, {van den Bosch}, {Verolme},
  \& {de Zeeuw}}]{vandeven2006}
{van de Ven} G., {van den Bosch} R.~C.~E., {Verolme} E.~K., {de Zeeuw} P.~T.,
  2006, \aap, 445, 513

\bibitem[{{van den Bergh}(1994)}]{vandenbergh1994}
{van den Bergh} S., 1994, \apj, 428, 617

\bibitem[{{Vanture} {et~al.}(2002){Vanture}, {Wallerstein}, \&
  {Suntzeff}}]{vanture2002}
{Vanture} A.~D., {Wallerstein} G., {Suntzeff} N.~B., 2002, \apj, 569, 984

\bibitem[{{Venn} \& {Hill}(2005)}]{venn2005}
{Venn} K.~A., {Hill} V., 2005, in IAU Symposium, Vol. 228, From Lithium to
  Uranium: Elemental Tracers of Early Cosmic Evolution, {Hill} V., {Fran{\c
  c}ois} P., {Primas} F., eds., pp. 513--518

\bibitem[{{Ventura} \& {D'Antona}(2005{\natexlab{a}})}]{ventura2005a}
{Ventura} P., {D'Antona} F., 2005{\natexlab{a}}, \aap, 431, 279

\bibitem[{{Ventura} \& {D'Antona}(2005{\natexlab{b}})}]{ventura2005b}
---, 2005{\natexlab{b}}, \aap, 439, 1075

\bibitem[{{Ventura} \& {D'Antona}(2005{\natexlab{c}})}]{ventura2005c}
---, 2005{\natexlab{c}}, \apjl, 635, L149

\bibitem[{{Villanova} {et~al.}(2007){Villanova}, {Piotto}, {King}, {Anderson},
  {Bedin}, {Gratton}, {Cassisi}, {Momany}, {Bellini}, {Cool}, {Recio-Blanco},
  \& {Renzini}}]{villanova2007}
{Villanova} S., {Piotto} G., {King} I.~R., {Anderson} J., {Bedin} L.~R.,
  {Gratton} R.~G., {Cassisi} S., {Momany} Y., {Bellini} A., {Cool} A.~M.,
  {Recio-Blanco} A., {Renzini} A., 2007, \apj, 663, 296

\bibitem[{{Woosley} \& {Weaver}(1995)}]{woosley1995}
{Woosley} S.~E., {Weaver} T.~A., 1995, \apjs, 101, 181

\end{thebibliography}
   
\label{lastpage}    
\end{document}